\documentclass{aa}

\usepackage{graphicx}
\usepackage{txfonts}
\usepackage{natbib,twoopt}
\usepackage[breaklinks=true]{hyperref} 
\bibpunct{(}{)}{;}{a}{}{,}             
\makeatletter
  \newcommandtwoopt{\citeads}[3][][]{\href{http://adsabs.harvard.edu/abs/#3}%
    {\def\hyper@linkstart##1##2{}%
     \let\hyper@linkend\@empty\citealp[#1][#2]{#3}}}
  \newcommandtwoopt{\citepads}[3][][]{\href{http://adsabs.harvard.edu/abs/#3}%
    {\def\hyper@linkstart##1##2{}%
     \let\hyper@linkend\@empty\citep[#1][#2]{#3}}}
  \newcommandtwoopt{\citetads}[3][][]{\href{http://adsabs.harvard.edu/abs/#3}%
    {\def\hyper@linkstart##1##2{}%
     \let\hyper@linkend\@empty\citet[#1][#2]{#3}}}
  \newcommandtwoopt{\citeyearads}[3][][]%
    {\href{http://adsabs.harvard.edu/abs/#3}
    {\def\hyper@linkstart##1##2{}%
     \let\hyper@linkend\@empty\citeyear[#1][#2]{#3}}}
\makeatother
\begin{document}

   \title{Spectroscopic classification and Gaia DR2 parallaxes of new nearby white dwarfs among selected blue proper motion stars\thanks{Based on observations with the 2.2\,m telescope of the Centro 
Astron\'omico Hispano-Aleman (CAHA), Calar Alto, Spain}}

   \subtitle{}

   \author{R.-D. Scholz
          \inst{1}
          \and
          H. Meusinger\inst{2,3}
          \and
          A. Schwope\inst{1}
          \and
          H. Jahrei{\ss}\inst{4}
          \and
          I. Pelisoli\inst{5,6}
          }

   \institute{Leibniz-Institut f\"ur Astrophysik Potsdam,
              An der Sternwarte 16, D--14482 Potsdam, Germany\\
              \email{rdscholz@aip.de,aschwope@aip.de}
         \and
             Th\"uringer Landessternwarte Tautenburg,
             Sternwarte 5, D--07778 Tautenburg, Germany\\
             \email{meus@tls-tautenburg.de}
         \and
             University Leipzig, Faculty of Physics and Geosciences,
             Linn\`estr. 5, D--04103 Leipzig, Germany
         \and
             Zentrum f\"ur Astronomie der Universit\"at Heidelberg,
             Astronomisches Rechen-Institut,
             M\"{o}nchhofstra\ss{}e 12-14, 69120 Heidelberg, Germany\\
             \email{hartmut@ari.uni-heidelberg.de}
         \and
             Instituto de F\'isica, 
             Universidade Federal do Rio Grande do Sul,
             91501-900, Porto-Alegre, RS, Brazil
         \and
             Institut f\"ur Physik und Astronomie, 
             Universit\"atsstandort Golm,
             Haus 28, Karl-Liebknecht-Str. 24/25, 14467 Potsdam, Germany\\
             \email{pelisoli@astro.physik.uni-potsdam.de}
             }

   \date{Received 22 June 2018; accepted 6 August 2018}


  \abstract
   {}
   {With our low-resolution spectroscopic observing program for
   selected blue proper motion stars, we tried to find new white dwarfs 
   (WDs) in the solar neighbourhood.}
   {We used the L{\'e}pine \& Shara Proper Motion (LSPM) catalogue
   with a lower proper motion limit of 150\,mas/yr and the
   Second US Naval Observatory CCD Astrograph Catalog (UCAC2)
   for proper motions down to about 90\,mas/yr. The LSPM and
   UCAC2 photometry was combined with Two Micron All Sky Survey (2MASS) 
   near-infrared (NIR) photometry. 
   Targets selected according to their 
   blue optical-to-NIR and NIR colours were observed mainly at Calar Alto.
   The spectra were classified by comparison with a large
   number of
   already known comparison objects, including WDs, 
   simultaneously observed 
   within our program. Gaia DR2 parallaxes and colours were used to
   confirm or reject spectroscopic WD candidates and to derive improved
   effective temperatures.
   }
   {We found ten new WDs at distances between 24.4\,pc and 79.8\,pc,
   including 
   six hot DA WDs:
   \object{GD 221} (DA2.0), 
   \object{HD 166435 B} (DA2.2), 
   \object{GD 277} (DA2.2),
   \object{2MASS J19293865+1117523} (DA2.4),   
   \object{2MASS J05280449+4105253} (DA3.6), and
   \object{2MASS J05005185-0930549} (DA4.2).
   The latter is 
   rather bright ($G\approx12.6$) and
   with its Gaia DR2 parallax of $\approx14$\,mas 
   it appears overluminous
   by about 3\,mag compared
   to the WD sequence in the Gaia DR2 colour-magnitude diagram.
   It may be the closest extremely low mass (ELM) WD
   to the Sun. 
   We further classified 
   \object{2MASS J07035743+2534184} as DB4.1. 
   With its
   distance of 25.6\,pc 
   it is the second nearest known representative of its class.
   With
   \object{GD 28} (DA6.1),
   \object{LP 740-47} (DA7.5), and
   \object{LSPM J1919+4527} (DC10.3)
   three additional cool WDs were found.
Gaia DR2 parallaxes showed us that four of
   our candidates but also two previously
   supposed WDs
   (\object{WD 1004+665} and \object{LSPM J1445+2527}) are in fact
   distant
   Galactic halo stars with high tangential velocities. Among our rejected
   WD candidates, we identified a
   bright ($G=13.4$\,mag)
   G-type carbon dwarf,
   \object{LSPM J0937+2803}, at a distance of 272\,pc.
   }
   {}


\titlerunning{Spectroscopic classification and Gaia DR2 parallaxes of new nearby WDs}
   \maketitle
%

\section{Introduction}
\label{Sect_intro}

The stellar content of the Solar neighbourhood provides important
information for the study of the stellar initial mass function,
Galactic structure and evolution. White dwarfs (WDs) play an important
role in our understanding of these fundamental astrophysical 
issues \citepads{2005ARA&A..43..247R,2014ApJ...791...92T}.
In particular, the cool WDs provide age estimates of the
Galactic disk \citepads{2015MNRAS.449.3966G,2017ApJ...837..162K}
and information on the baryonic 
dark matter in the Galaxy 
\citepads{2001ApJ...559..942R,2002A&A...395..779M,2003MNRAS.339..817F}.
The nearest WDs are also excellent targets for divers methods
to search for extrasolar
planets \citepads{2010ApJ...708..411K,2015A&A...579L...8X}.

As the general census of nearby stars has been slowly improved in
the pre-Gaia era \citepads[see e.g.][]{2018AJ....155..265H}
the sample of known WDs in the solar neighbourhood experienced a
similar evolution. 
While the WD census was believed to be complete within
13\,pc \citepads{2016MNRAS.462.2295H}, during the last ten years
many efforts were undertaken
to identify more WDs in the extended solar neighbourhood (e.g. for
the 25\,pc and 40\,pc samples) by measuring their parallaxes
\citepads{2009AJ....137.4547S,2017AJ....154...32S} and using
spectroscopic surveys
\citepads{2011ApJ...743..138G,2012MNRAS.425.1394K, 
2013AJ....145..136L,2014AJ....147..129S}.
The physical properties of nearby WDs were investigated in more
detail by \citetads{2012ApJS..199...29G}, \citetads{2015ApJS..219...19L},
\citetads{2017MNRAS.465.2849T}, \citetads{2017MNRAS.467.4970H}, and
\citetads{2017AJ....154..118S}. The overall number of spectroscopically
classified WDs
(including more distant objects) was increased fivefold with the
catalogues of \citetads{2013ApJS..204....5K} 
and \citetads{2015MNRAS.446.4078K,2016MNRAS.455.3413K},
based on spectroscopic data from the
Sloan Digital Sky Survey \citepads[SDSS;][]{2000AJ....120.1579Y}.

Despite of all the above mentioned efforts,
even the generally assumed completeness of the 13\,pc WD 
sample was shown to be not correct, as the recent discovery of 
the cool WD companion of a new nearby red dwarf
star \object{TYC 3980-1081-1 B} \citepads{2018A&A...613A..26S}
at a distance of 8.3\,pc
indicated.
This discovery was based on astrometric data from
Gaia DR1 \citepads{2016A&A...595A...4L},
the 5th United States Naval Observatory CCD Astrograph 
Catalog \citepads[UCAC5;][]{2017AJ....153..166Z},
and the U.S. Naval Observatory Robotic Astrometric Telescope
\citepads[URAT;][]{2015AJ....150..101Z}
Parallax Catalog \citepads[UPC;][]{2016yCat.1333....0F}.
The WD status of \object{TYC 3980-1081-1 B} 
is now confirmed by its Gaia DR2 
\citepads{2018arXiv180409365G} parallax of 118.12$\pm$0.02\,mas.

Most searches for previously unrecognised nearby stars in the pre-Gaia 
era investigated targets selected from combined colour- and proper 
motion surveys. In 2008, we started a spectroscopic follow-up
programme to search for missing WDs in the solar neighbourhood
in a sample of blue proper motion stars. The L{\'e}pine \& Shara
Proper Motion \citepads[LSPM; ][]{2005AJ....129.1483L} catalogue
of the northern sky and the Second US Naval Observatory CCD Astrograph
Catalog \citepads[UCAC2; ][]{2004AJ....127.3043Z}
covering the sky area from $-90\degr$ to $+40\degr$ declination
(going up to $+52\degr$ in some areas), and the 
Two Micron All Sky Survey \citepads[2MASS;][]{2006AJ....131.1163S}
providing accurate near-infrared (NIR) photometry
served as our main input data with respect to proper motions
and photometry.
The results of our WD survey were not published so far, except 
for one new WD (\object{HD 166435 B} = \object{CA376}) that was 
independently discovered in a search for common proper motion WD
companions of known nearby stars \citepads{2018A&A...613A..26S}.

This paper is organised as follows:
In Sect.~\ref{Sect_sample} we describe the selection of our target
stars for the spectroscopic observations, which were mainly carried
out at the Calar Alto observatory (Sect.~\ref{Sect_spec}). 
Sect.~\ref{Sect_class} deals with our spectroscopic classification of 
WD candidates of different classes, 
whereas Sects.~\ref{Sect_Plxg} and \ref{Sect_BP_RP}
show how well our results are confirmed by Gaia DR2 parallaxes
and colours. Finally, we review previous investigations
on our confirmed and rejected WDs and briefly discuss the properties
of the most interesting objects in Sect.~\ref{Sect_discuss}.

%

\section{Sample of blue proper motion stars}
\label{Sect_sample}

For the nearest stars in the 10\,pc sample, the Research Consortium
on Nearby Stars (RECONS) had reported statistics for the time from 2000 
to 2012\footnote{http://www.chara.gsu.edu/RECONS/}. According to this,
the number of WDs increased from 18 to 20, whereas that of other stars
(and brown dwarfs) rose from 273 to 337. It has long been known that
in general
the nearest known WDs have extraordinary high proper motions compared 
to 
the other nearest stars. 
Using SIMBAD, \citetads{2015A&A...574A..96S} 
found 84\% of 
the WDs within 10\,pc
have proper motions 
larger than 
1000\,mas/yr, whereas 
the fraction for the other stars 
was only 49\%. Motivated by the discrepancy between the numbers 
of expected and known WDs, we launched a project to
find new WD candidates among stars with moderately high proper 
motions as surveyed in the LSPM catalogue, which has a lower proper
motion limit of
150\,mas/yr. In addition, we also selected candidates from the UCAC2 
catalogue, which was not the result of a dedicated high proper motion 
survey but contained relatively accurate proper motions of all stars
to about 16th magnitude in a large portion of the sky, to push 
the proper motion limit for our candidates to even lower values 
around 100\,mas/yr.

We investigated the distribution of 
all known WDs within 15\,pc compared 
to all LSPM stars in an optical-to-NIR colour-magnitude diagram 
(CMD) of
$J$ vs. $RF-K_{\rm s}$, with the NIR magnitudes $J$ and $K_{\rm s}$ from 2MASS
and the photographic red magnitude $RF$ given in the LSPM. We also
looked at the pure NIR CMD 
of $J$ vs. $J-K_{\rm s}$, 
providing more accurate colours than the photographically
determined LSPM ones.
Our colour and magnitude
cuts for the selection of new WD candidates corresponded to the
observed areas occupied by the known nearby WDs in both of 
these two CMDs. We decided not to use
a reduced proper motion diagram for 
the selection of targets, since we were especially
interested to find 
new WDs
among the objects with only
moderately large proper motions (and small tangential velocities). 
So our LSPM WD candidates were selected by the following criteria:
\begin{itemize}
\item proper motion $>$150\,mas/yr 
\item $K_{\rm s}<15$\,mag
\item $RF-K_{\rm s}<+1.75$\,mag and $J-K_{\rm s}<+0.5$\,mag if $J>12.5$\,mag
\item $RF-K_{\rm s}<+0.75$\,mag and $J-K_{\rm s}<+0.3$\,mag if $J<12.5$\,mag
\end{itemize}
At the beginning of our survey, in 2008, we found
that according to SIMBAD bibliography
one third of our 
approximately
600 initial LSPM candidates
were
already known either as 
WDs \citepads[][with updates until 2008]{1999ApJS..121....1M}
or as Galactic thick disk and halo stars (including uncertain
candidates). In particular,
all targets with proper motions 
larger than
500\,mas/yr turned out to be known WDs. Meanwhile
these numbers have increased considerably, 
because new WDs
were added mainly thanks to the efforts 
of \citetads{2013AJ....145..136L} and \citetads{2015ApJS..219...19L}. 
They exploited 
not only
the LSPM 
catalogue but also its 
unpublished 
extension to a lower
proper motion limit of 40\,mas/yr and used a combination of
colour-magnitude and reduced proper motion diagrams in their
selection of WD candidates for spectroscopic follow-up observations. 
Interestingly, 19 out of 54 known WD comparison
objects used in our classification (Sect.~\ref{Sect_class}) 
were spectroscopically classified as WDs only
after 2009,
i.e. after our spectroscopic observing runs.

We selected additional targets with smaller proper motions from the 
UCAC2 catalogue if they fulfilled the following criteria:
\begin{itemize}
\item proper motion $>$50\,mas/yr
\item $UCmag>10$\,mag
\item $UCmag-K_{\rm s}<+1.1$\,mag and $J-K_{\rm s}<0.5$\,mag
\item proper motion confirmed with multi-epoch finder charts
\end{itemize}
The UCAC2 magnitudes $UCmag$, with a bandpass between the $V$ and $R$
bands, are probably of higher quality than photographic magnitudes 
but 
according to the catalogue description were
often observed under non-photometric conditions, and their absolute
and relative errors are given 
to 0.3\,mag and 0.1\,mag.
As the UCAC2 contains spurious 
high
proper motion objects, we checked them
all by inspecting their multi-epoch positions using the finder chart 
tool\footnote{http://irsa.ipac.caltech.edu/applications/FinderChart/}.
By this procedure, the number of UCAC2 candidates 
could be significantly reduced to about 450 objects
(including known WDs as well as other known stars according to SIMBAD),
especially in crowded regions of the Galactic plane.
Compared to the LSPM WD candidates, the UCAC2 WD candidates were 
given lower priorities the smaller their proper motion is. Effectively,
they were only observed if their proper motions were larger than
about 90\,mas/yr and when no other suitable LSPM candidates were 
available.

\section{Low-resolution spectroscopic observations}
\label{Sect_spec}

The majority of the known WDs typically show relatively blue colours.
Cool WDs, on the other hand, have only moderately blue or even red 
optical colours \citepads{1999ApJ...520..680H}, which we took into
account with our colour cuts described in Sect.~\ref{Sect_sample}.
By doing so, we expected a substantial contamination of our
target sample by high-velocity F-K subdwarfs
located at larger distances from the Sun. The adequate method to 
distinguish WDs from contaminating subdwarfs in a large sample of
candidates is low-resolution spectroscopy. This remains true even though
observations with low signal-to-noise (S/N) and reaching not far enough to
the blue part of the 
spectrum, where typical atomic (e.g. lines of Na and Ca) and 
molecular (e.g. the MgH band at 5200\,{\AA}) absorption features 
of subdwarfs appear, 
can be misleading
in the classification of cool DC WDs (Sect.~\ref{Sect_coolDC}), 
expected to show featureless spectra, and 
subdwarfs \citepads[compare][]{2004A&A...425..519S,2005ApJ...627L..41F}.

Considering low-resolution stellar spectra dominated by 
strong Balmer lines, it may also become difficult to distinguish 
relatively hot DA WDs from 
so-called sdA stars \citepads{2018MNRAS.475.2480P}. The latter are
thought to represent various byproducts of binary evolution and 
include the class of relatively rare extremely low mass (ELM) WDs
\citepads{2016ApJ...818..155B}. As there are now Gaia parallaxes
available, we can estimate absolute magnitudes of such ELM candidates
and compare them with those of normal WDs (see Sect.~\ref{Sect_Plxg}).

For the vast majority ($\approx 90$\%) of our targets, 
low-resolution spectra were obtained with the Calar Alto focal reducer 
and faint object spectrograph (CAFOS) at the 2.2 m telescope of the Centro 
Astron\'omico Hispano-Aleman (CAHA), Calar Alto, Spain. The blue grism 
B200 was used in combination with the Site-1d CCD resulting in a pixel 
size corresponding to 0.53 arcsec and a dispersion of 4.7\,{\AA} per pixel, 
or about 15\,{\AA} FWHM for a slit width equal to the typical seeing of 
1-2 arcsec.  For a small subsample of our targets we used the Nasmyth 
Focal Reducer Spetrograph (NASPEC) at the 2\,m Tautenburg telescope 
equipped with a Site-T4a CCD chip with a pixel size corresponding 
to 0.53 arcsec. The V-200 grism was used, which yields a wavelength 
coverage from 4000\,{\AA} to 8500\,{\AA} and a dispersion of 3.4\,{\AA} 
per pixel, or 12\,{\AA} FWHM for a typical slit width of 1 arcsec.

The results of similar spectroscopic observations with NASPEC and CAFOS
of three previously suspected WDs were presented by 
\citetads{2008A&A...484..575J}. All three stars turned out to
be high-velocity F- to K-type stars. Interestingly, the low 
S/N NASPEC spectrum of one of these objects (\object{GJ 2091})
first 
pointed
at a possible cool WD nature, whereas the CAFOS spectrum 
\citepads[Fig.2 in][]{2008A&A...484..575J} allowed its classification 
as a distant halo star with a very large tangential
velocity $>$500\,km/s (now confirmed with Gaia DR2 as 603\,km/s)
and a measured radial velocity of about 270\,km/s.
In comparison to NASPEC the CAFOS spectroscopy
had not only
the advantage of usually better observing conditions
but also of 
its extension 
to the blue spectral range below 4000\,{\AA}
including the Ca H and K lines, critical for the classification of 
non-WDs (see Sect.~\ref{Sect_rejected}).

The CAFOS observations were carried out in visitor mode
(5 nights in August 2008 and 5 nights in April 2009) by HM and RDS
and in service mode (4 nights in December 2008, 1 night in April 2009). 
The NASPEC observations were made by HM (5 nights in February 2008).
Exposure times between 2 and 20 minutes (with multiple observations
for some of our faintest targets) were sufficient to reach
a S/N of at least 10-20
for our targets in the typical magnitude range
$12<RF<16$\,mag.

The data reduction was performed under ESO MIDAS and 
followed the standard procedure for long-slit spectroscopy.
We used two different software packages, developed for CAFOS and 
NASPEC spectra. The main differences are in the 
extraction of one-dimensional spectra and in the flux and wavelength 
calibration. CAFOS shows substantial distortions of the plate scale 
near the edges of the field, which is not the case for NASPEC.
The CAFOS reduction uses the optimum extraction algorithm by 
\citetads{1986PASP...98..609H}, whereas a 
simple extraction method was implemented in 
the NASPEC version. According to our experience, the results from 
both methods show a good agreement for most stellar 
spectra \citepads{2005A&A...442..211S}. The Horne algorithm provides a 
better S/N for the faintest stars, but is more sensitive 
to variations of the background over the field. Wavelength calibration 
was done with calibration lamp spectra in the CAFOS version and with 
the night-sky lines \citepads{1992PASP..104...76O} present in the 
target spectra for the NASPEC observations.  In both versions, a 
relative flux calibration is implemented that corrects mainly 
for the spectral sensitivity of the camera. More precise flux 
calibration was 
obtained 
by means of simultaneously observed 
spectrophotometric standard stars afterwards.

All of the WDs and WD candidates described in this paper 
were observed with CAFOS. On the other hand, there was no new 
WD candidate found among the NASPEC spectra. In Sect.~\ref{Sect_class}, 
we show therefore only CAFOS spectra, which were reduced with the CAFOS 
reduction package.

As by far not all our candidates could be observed because of bad weather
conditions and visibility problems during our scheduled observing runs, 
our final spectroscopic data set consists of 490 spectra 
for 410 objects. Multiple observations were carried out for some 
of the known comparison objects, for some objects that we selected for
observations at both Calar Alto and Tautenburg, and for our faintest 
targets. The 490 spectra 
finally obtained were sorted by their 
target coordinates and 
observing times.
All objects were named by their first occurence in 
that list, 
with names between ``CA001'' and ``CA490''. These 
designations 
are used throughout
this paper. 
Additional designations
from SIMBAD, in particular of previously 
known WDs, are given in the corresponding figure captions and
in the text. 
In Tables~\ref{Tab_10WDs} and \ref{Tab_6nonWDs}, we provide 
also alternative designations for all new and rejected WDs.

   \begin{figure*}
   \centering
   \includegraphics[width=15.0cm]{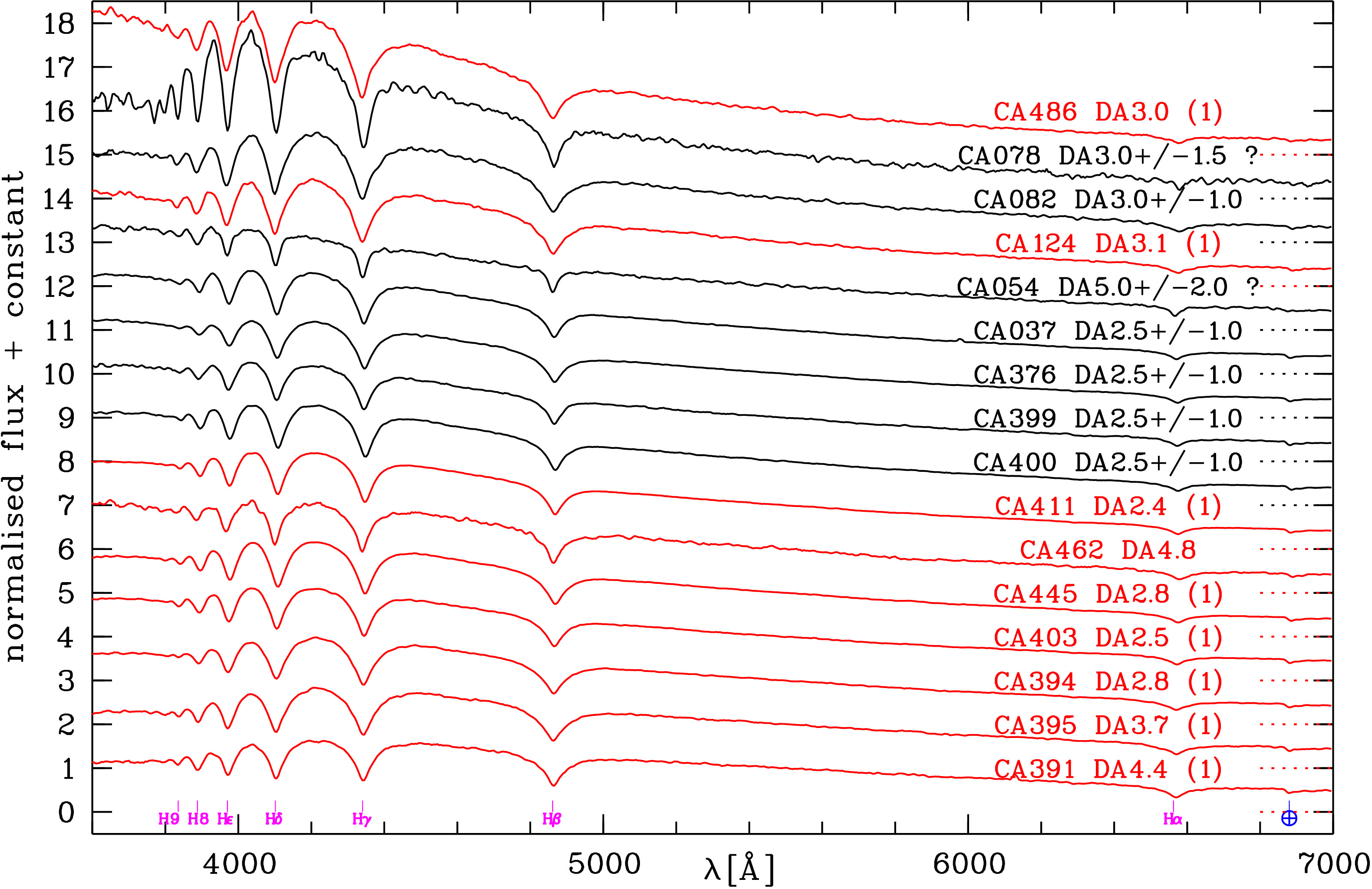}
      \caption{CAFOS spectra of known (red) and new (black) hot DA WDs.
              The spectral types of known objects,
              \object{CA486} = \object{WD 2329+407},
              \object{CA124} = \object{WD 0809+177},
              \object{CA411} = \object{WD 2032+248},
              \object{CA462} = \object{WD 2246+223},
              \object{CA445} = \object{WD 2149+021},
              \object{CA403} = \object{WD 1943+163},
              \object{CA394} = \object{WD 1911+536},
              \object{CA395} = \object{WD 1911+135},
              \object{CA391} = \object{GD 532},
              have uncertainties
              below $\pm$0.1 subtypes according to reference (1) =
              \citetads{2015ApJS..219...19L}. The spectra are roughly
              sorted by the slope of their continuum. The locations of
              the Balmer lines and of a terrestrial absorption band are
              marked on the bottom.
              }
         \label{Fig_hotDA}
   \end{figure*}

\section{WD classification}
\label{Sect_class}

Our spectroscopic classification of new WD candidates was mainly 
based on the direct comparison with known WDs, which were observed
with the same instrument. 
We applied the WD spectral classification system of 
\citetads{1983ApJ...269..253S} illustrated with an atlas of optical 
WD spectra \citepads{1993PASP..105..761W}. All our 
spectra were normalised to their
flux in the wavelength interval 5400-5600\,{\AA} and visually
compared with each other, including many different classes of
known non-WD objects (sdO, sdB, AFG-type stars, low-metallicity
F and G subdwarfs). 
The newly
classified non-WD objects will be published in a separate paper
(Scholz et al., in preparation), the various new WD types found
are described in the following 
subsections. In most cases our spectra are of sufficient
quality for the classification.
However, 
some spectra had only
a low S/N ratio 
or no clear matching with the spectra of
known comparison objects so that we considered those objects 
only as
WD candidates, to which we assigned spectral types with question marks.
The results of our initial spectroscopic classification described 
in this section are shown in Tables~\ref{Tab_10WDs} and \ref{Tab_6nonWDs},
where we also list our final and more accurate spectral types determined
with the aid of
Gaia DR2 colours (see Sect.~\ref{Sect_BP_RP}).

\subsection{New hot DA WDs and candidates}
\label{Sect_hotDA}

Typical hot WDs in the solar neighbourhood, with hydrogen atmospheres 
and effective temperatures between
10\,000\,K and 25\,000\,K (spectral types from $\approx$DA5.0 to
$\approx$DA2.0)
generally show relatively broad Balmer lines and a blue continuum
in their spectra. Compared to A-type stars, the hot DA WDs
exhibit a less prominent decline at the Balmer jump, as
their higher-level Balmer lines become weaker
due to 
the blanketing effect \citepads{2013AJ....146...34Z}.
However, as also mentioned by \citetads{2013AJ....146...34Z},
who studied DA WDs in the Large Sky Area Multi-Object Fiber 
Spectroscopic Telescope (LAMOST) survey with slightly higher
spectral resolution of $\sim$1500, the Balmer line widths 
of hot DA WDs of lower surface gravity are narrower
and may appear similar to those of A-type stars. In a similar 
way as done by  \citetads{2013AJ....146...34Z},
we classified objects with blue spectra showing the Ca K line 
in addition to the Balmer lines as A-type (or early F-type) stars.
By doing so, we may in principle have excluded some relatively 
hot DAZ WDs like \object{GD 362} \citepads{2004ApJ...617L..57G}.
However, as the Gaia parallaxes have shown (Sect.~\ref{Sect_Plxg}), 
we do not miss any WD in our classification.

\begin{figure*}
   \sidecaption
   \includegraphics[width=10.9cm]{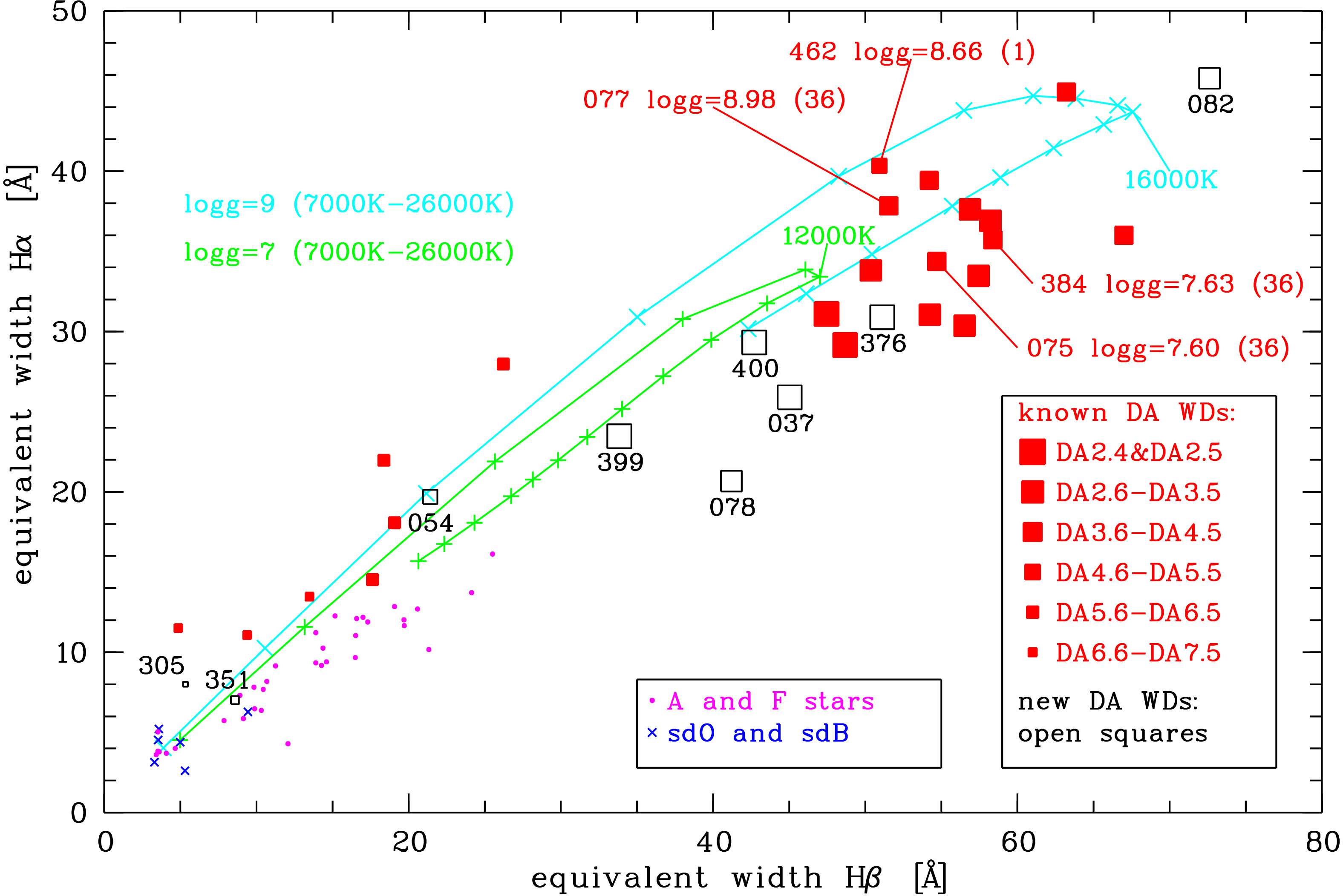}
      \caption{Equivalent widths of H$\beta$
               and H$\alpha$ lines measured in the CAFOS
               spectra of DA WDs and other objects with
               dominant Balmer lines. The classification
               of known DA WDs with typical
               errors of less than about 0.1 subtypes 
               was taken from
               (1) = \citetads{2015ApJS..219...19L}, 
               (2) = \citetads{2012MNRAS.425.1394K}, and
               (36) = \citetads{2011ApJ...743..138G}.
               Their gravities, given by these authors,
               varied between $7.6<\log{g}<9.0$. The objects
               with the two lowest and highest gravities,
              \object{CA462} = \object{WD 2246+223},
              \object{CA384} = \object{WD 1824+040}, 
              \object{CA075} = \object{WD 0416+701},
              \object{CA077} = \object{WD 0457-004},
               are marked.
               New DA WDs (and candidates) 
               from Figs.~\ref{Fig_hotDA}
               and \ref{Fig_coolDA} are also labelled by their 
               CA numbers. The plus signs and crosses 
               connected with solid lines show
               the measurements of model spectra with $T_{\rm eff}$ from
               7000\,K to 26\,000\,K (below 20\,000\,K in steps of 
               1000\,K, above 20\,000\,K in steps of 2000\,K) 
               with gravities of $\log{g}=7$
               and $\log{g}=9$.
              }
         \label{Fig_HaHbEW}
   \end{figure*}

\begin{figure*}
   \sidecaption
   \includegraphics[width=10.9cm]{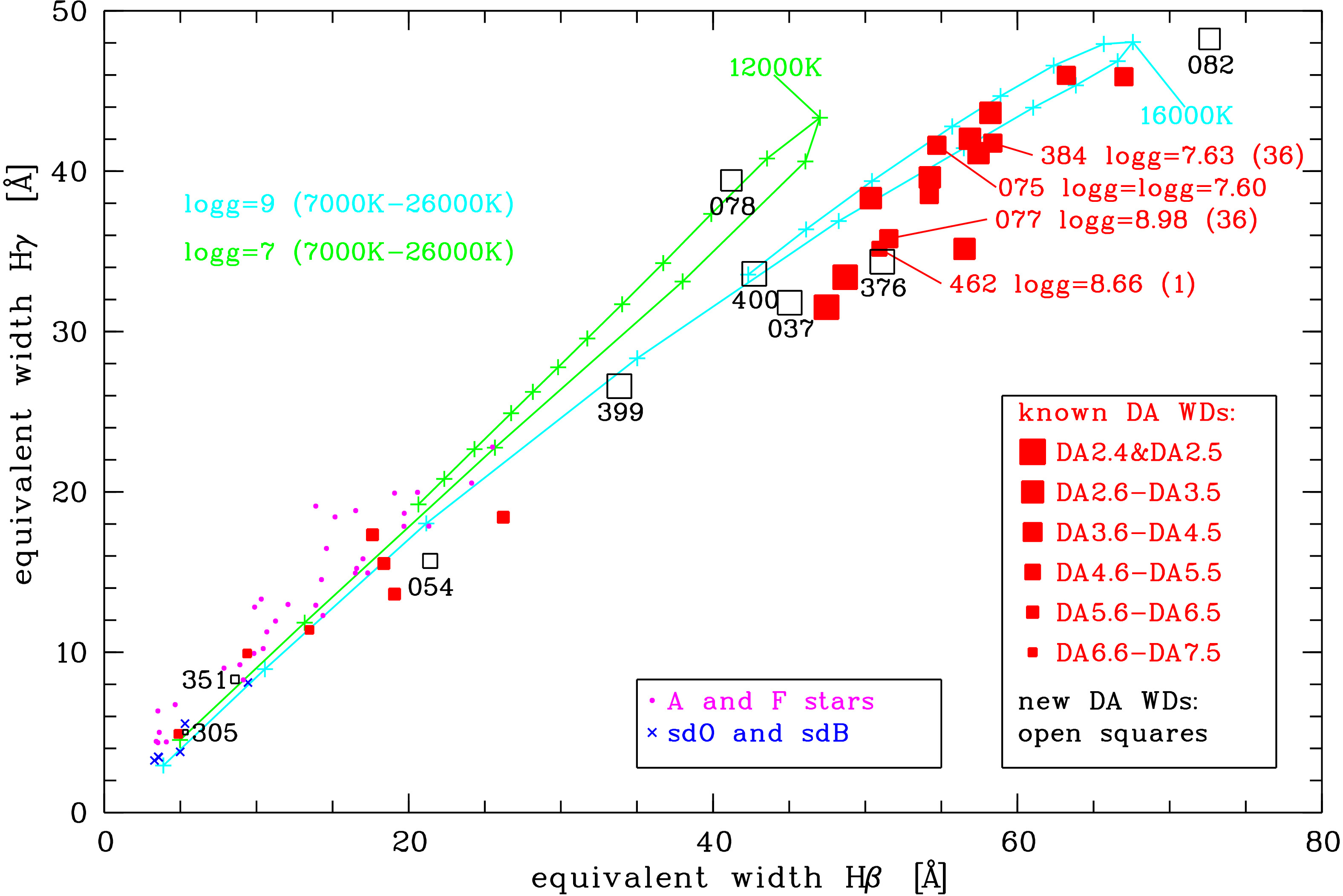}
      \caption{Same as Fig.~\ref{Fig_HaHbEW} 
               for equivalent 
               widths of H$\beta$ and H$\gamma$ lines.
              }
         \label{Fig_HbHgEW}
   \end{figure*}

As DA WDs should not show higher-level Balmer lines in low-resolution
spectra \citepads{1992ApJS...78..409B}, we used their appearance
as an additional criterion to classify blue objects with Balmer lines
only. All non-DA objects were excluded from further analysis in this 
study, but are included 
in Figs.~\ref{Fig_HaHbEW} and \ref{Fig_HbHgEW}, 
showing 
H$\alpha$, H$\beta$, and H$\gamma$
line width measurements, for comparison.

The spectra of our new and known hot DA WDs, all observed with
CAFOS, are presented in Fig.~\ref{Fig_hotDA}. Here we show only those 
of the known hot DA WDs from our sample that were included in the
current census of northern WDs 
within 40\,pc \citepads{2015ApJS..219...19L} and had consistent
spectral types from other sources, too. As one can see, there is only
a weak correlation of their spectral types with the slopes of 
the continuum in our spectra. As the reference spectral types are,
according to the errors in effective temperature ($\lesssim$150-300\,K),
all accurate to within $\pm$0.1 subtypes, we assumed that the spectral 
slopes at the blue end of our spectra are affected by uncertainties in 
the flux calibration. However, 
we note that 
the strongest outlier
in the sequence of known objects shown in Fig.~\ref{Fig_hotDA},
\object{CA462} (= \object{WD 2246+223})
with a spectral type of DA4.8, 
appears 
as blue as the neighbouring objects of much earlier (DA2.4-DA2.8) types
and has 
similar line widths as these objects (see below 
in Figs.~\ref{Fig_HaHbEW} and \ref{Fig_HbHgEW}).

Our spectral classification of new hot DA WDs relied on the closest
matches with the spectra of known comparison objects. Four spectra,
\object{CA037}, \object{CA376}, \object{CA399}, and \object{CA400},
are very similar to our hottest comparison DA WD, 
\object{CA411} (= \object{WD 2032+248}; DA2.4).
One object, 
\object{CA082},
resembles the DA3.1 WD \object{CA124} (= \object{WD 0809+177}) 
(plotted next
to each other in Fig.~\ref{Fig_hotDA}). Because of
the weak correlation of known spectral types with the slopes of our
spectra, we assigned spectral types in steps of 0.5 subtypes and
uncertainties of 1.0 subtypes to these new objects. 
For
the noisy spectrum of \object{CA078}, which has the closest match to
the 
DA3.0 spectrum of \object{CA486} (= \object{WD 2329+407}), 
shown on top of 
Fig.~\ref{Fig_hotDA}, we conservatively assigned an uncertainty
of 1.5 subtypes. 
As a special case with an even larger
uncertainty of 2.0 spectral subtypes we consider the relatively
blue spectrum, probably caused by an error in the flux calibration, 
of \object{CA054}, which we classified as DA5.0
(also shown in Fig.~\ref{Fig_coolDA}) because of its relatively 
narrow Balmer lines 
(see also Figs.~\ref{Fig_HaHbEW} and \ref{Fig_HbHgEW}).

We measured the equivalent widths of the 
H$\alpha$, H$\beta$, and H$\gamma$ lines 
in our low-resolution spectra of all known <DA7.5 WDs 
and of the new hot and cool DA WDs (Sect.~\ref{Sect_coolDA})
in our sample. In addition,
we measured the same three lines in the non-WD spectra
classified by us as 
A-type (or early F-type) stars (see above) and of known 
sdO and sdB among our comparison objects. The results are
presented 
in Figs.~\ref{Fig_HaHbEW} and \ref{Fig_HbHgEW}, 
where one can see a relatively clear 
separation of three groups of objects: hot DA WDs, cool DA WDs,
and non-WDs, 
in particular for H$\alpha$ as a function of H$\beta$.
The non-WDs reach equivalent widths of H$\beta$
up to about 25\,{\AA}, 
but show systematically smaller 
equivalent widths of H$\alpha$ than the cool DA WDs. 
The separation between cool DA WDs and non-WDs is less 
obvious 
in the H$\gamma$ vs. H$\beta$ diagram. 
The hot DA WDs exhibit large equivalent widths 
for all three lines,
where the maximum values are reached for types 
around DA4, which are known to have the strongest Balmer lines
at normal gravities of $\log{g}\approx8$ \citepads{1993PASP..105..761W}.

The four new hot DA WDs, \object{CA037}, \object{CA376}, 
\object{CA399}, and \object{CA400}, 
which were
all classified by us as DA2.5$\pm$1.0
based on the closest match of their spectra with that of
the known DA2.4 WD \object{CA411} (= \object{WD 2032+248}),
have
equivalent widths of 23$\lesssim$H$\alpha$$\lesssim$31\,{\AA}, 
33$\lesssim$H$\beta$$\lesssim$52\,{\AA}, and
26$\lesssim$H$\gamma$$\lesssim$35\,{\AA}.
These values are similar or slightly smaller compared to those of
the two known DA2.4 and DA2.5 WDs shown 
in Figs.~\ref{Fig_HaHbEW} and \ref{Fig_HbHgEW}.
This is a first indication that the four new DA2.5 WDs
may be even slightly hotter than estimated from our initial
classification.

One of the new hot DA WDs, \object{CA376} was independently discovered
as a wide companion to 
the well-known nearby star \object{HD 166435} and
already spectroscopically classified as DA2.2$\pm$0.2
by \citetads{2018A&A...613A..26S} based on the
comparison with only two known WDs, \object{CA411}
(= \object{WD 2032+248}; DA2.4)
and \object{CA445} (= \object{WD 2149+021}; DA2.8). This comparison 
made use 
of both the spectral slopes and the line widths of H$\beta$ and H$\alpha$.
Our new spectroscopic classification of \object{CA376}
as DA2.5$\pm$1.0, taking into account the observed spread in the
spectral slopes of a larger number of comparison objects, is
more conservative but consistent with the former classification.

\subsection{Comparison with model DA WD spectra}
\label{Sect_modelspec}

To check our spectral classification of DA WDs, in particular of the
hot DA WDs classified in Sect.~\ref{Sect_hotDA}, we also
compared our observed data with WD model atmosphere spectra
kindly made available by D.~Koester \citepads{2010MmSAI..81..921K}.
The spectra are for a pure
hydrogen atmospheric composition. Balmer lines in the models were
calculated with the modified  Stark broadening profiles
of \citetads{2009ApJ...696.1755T}
kindly made available by these authors to Koester. The models cover the
effective temperature range between 6\,000\,K and 100\,000\,K, and
a range in gravity from $\log{g}=5.00$ to $\log{g}=9.00$ with
a step size of 0.25 dex.
Due to the low S/N of our obtained spectra, we opted by comparing only 
equivalent widths to the model spectra, rather than doing a fit.
The spectra were normalised with low-order polynomials to selected
continuum points, and the equivalent widths of the Balmer lines (H$\alpha$,
H$\beta$, H$\gamma$) were measured with the INTEG/LINE command
within ESO/MIDAS for both model and observed spectra.

As expected, the model spectra show a general
trend to smaller equivalent widths with smaller gravity.
However, there are only weak gravity-dependent trends seen for the 
known DA WDs in our sample, whereas the noise-free model spectra
show clear changes in the equivalent widths, according to which e.g.
the turnoff point caused by the maximum of the Balmer lines
changes from about 16\,000\,K ($\approx$DA3.2)
at $\log{g}=9$ to about 12\,000\,K ($\approx$DA4.2) at $\log{g}=7$.

Interestingly, our two DA3.0 WDs \object{CA078} and \object{CA082}
are located at two edges of the distributions shown 
in both Figs.~\ref{Fig_HaHbEW} and \ref{Fig_HbHgEW}.
As the stronger H$\beta$ and H$\gamma$ lines in our hot DA spectra are 
generally better measured than the H$\alpha$ line, we see less scatter
of the observed data around the model data 
in Fig.~\ref{Fig_HbHgEW}. 
In that respect the location of \object{CA078}
as a clear outlier, falling in the region of the low-gravity
model curve, is remarkable. We will further discuss the possible low 
gravity status of this uncertain DA3.0$\pm$1.5 candidate 
in Sect.~\ref{Sect_CA078}. 
 
The better classified DA3.0$\pm$1.0 \object{CA082} lies 
close to the turnoff points for the $\log{g}=9$ model curves 
in both Figs.~\ref{Fig_HaHbEW} and \ref{Fig_HbHgEW}, 
in good agreement with
the effective temperature 
of about 16\,800\,K.
Finally, our very uncertain classification
of \object{CA054} as DA5.0$\pm$2.0 is supported by the
location of this object in the 
central regions of
both Figs.~\ref{Fig_HaHbEW} and \ref{Fig_HbHgEW} 
at the edges of the areas occupied by hot DA WDs, cool DA WDs,
and non-WDs.

\subsection{New cool DA WDs and candidates}
\label{Sect_coolDA}

\begin{figure*}
   \centering
   \includegraphics[width=15.0cm]{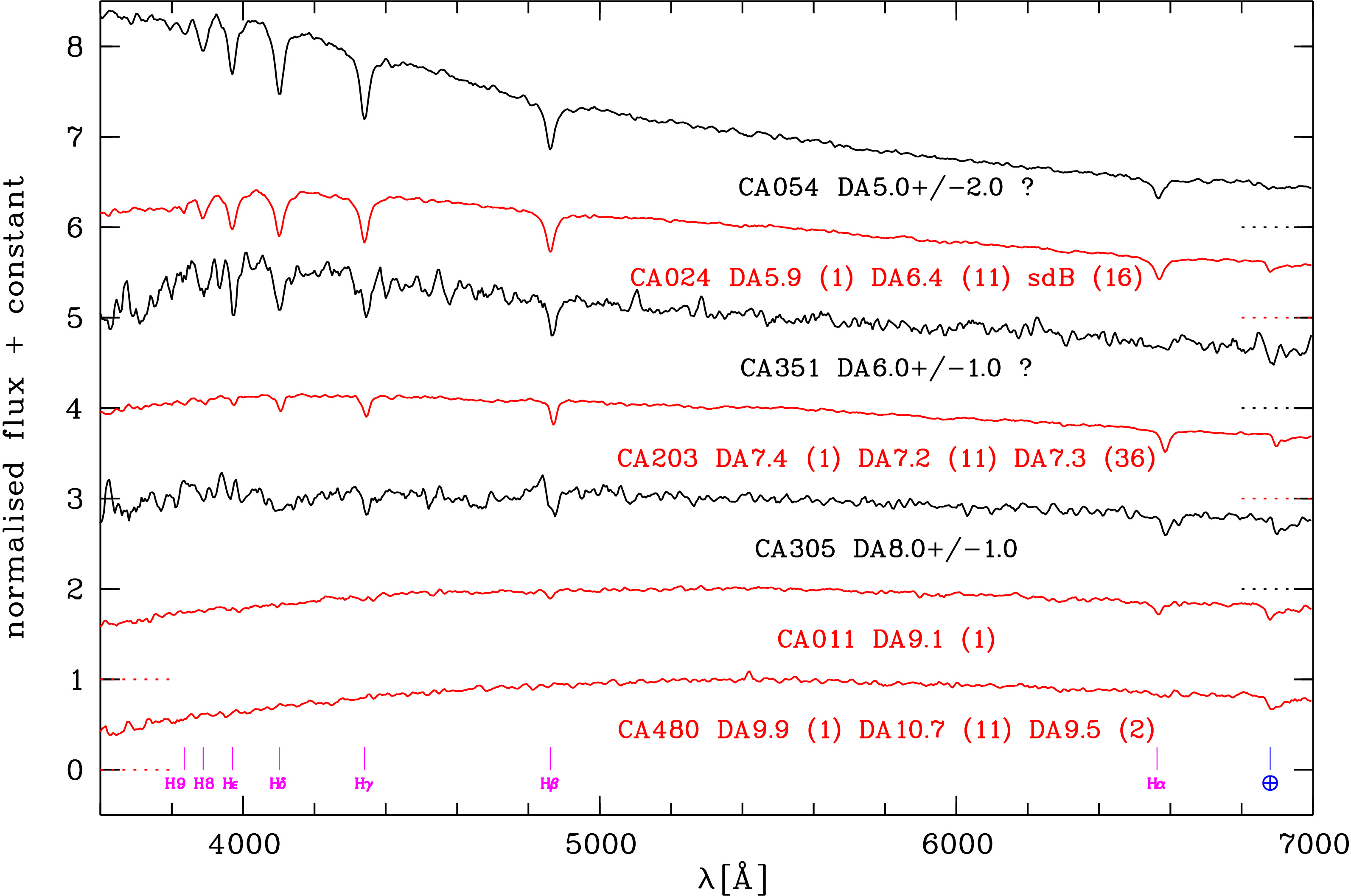}
      \caption{CAFOS spectra of known (red) and new (black) cool DA WDs.
              The spectral types of known objects,
              \object{CA024} = \object{WD 0101+048},
              \object{CA203} = \object{WD 1019+637},
              \object{CA011} = \object{WD 0025+054},
              \object{CA480} = \object{WD 2322+137},
              with typical
              errors of less than about 0.1 subtypes
              are taken from the following references: 
              (1) = \citetads{2015ApJS..219...19L},
              (11) = \citetads{2014AJ....147..129S}, 
              (16) = \citetads{2013A&A...551A..31D}, and
              (36) = \citetads{2011ApJ...743..138G}. 
              The spectra are roughly
              sorted by the slope of their continuum. The locations of
              the Balmer lines and of a terrestrial absorption band are
              marked on the bottom.
              }
         \label{Fig_coolDA}
   \end{figure*}

With decreasing effective temperatures the Balmer lines of DA WDs become 
sharper and weaker. As mentioned by \citetads{1993PASP..105..761W},
at low S/N ratio and with
the lowest-temperature DA WDs (beyond DA9) only H$\alpha$ remains
visible, and it becomes harder to distinguish between cool DA and
DC WDs (see Sect.~\ref{Sect_coolDC}). 
This can be seen in some of our spectra
shown in Figs.~\ref{Fig_coolDA} and \ref{Fig_coolDC}.

In Fig.~\ref{Fig_coolDA} we show our cool DA WD candidates
together with an representative sample of known DA5-DA10 WDs. 
The spectrum of \object{CA054} on top
was already discussed
in Sects.~\ref{Sect_hotDA} and \ref{Sect_modelspec}
and shown in Fig.\ref{Fig_hotDA}.
The Balmer line widths of this object look similar to those
of the known DA5.9 WD \object{CA024} (= \object{WD 0101+048}), 
whereas the continuum
of \object{CA054} is much bluer. The relatively noisy spectrum
of \object{CA351} was classified by us only as a weak DA6 candidate
(with question mark), because H$\alpha$ is not clearly visible, 
whereas the higher-level Balmer lines appear stronger. The spectrum 
of \object{CA305} shows the opposite trend of weakening higher-level
Balmer lines, also seen for the 
coolest known 
DA WDs at the bottom
of Fig.~\ref{Fig_coolDA}. Taking into account the slightly different
spectral classification of the comparison objects and the moderate
trend towards a bluer continuum with decreasing temperature, 
we assigned again uncertainties of 1.0 spectral subtypes
to our new cool DA candidates.

The classification of
\object{CA024} (= \object{WD 0101+048})
as sdB (see Fig.~\ref{Fig_coolDA})
seems to be a mismatch in SIMBAD. \citetads{2013A&A...551A..31D}
used the wrong name PG~0101+040 for the sdB \object{PG 0101+039},
and SIMBAD apparently by mistake associated this sdB to
\object{PG 0101+048}, which is in fact the known DA WD \object{CA024}
as confirmed with the first two and more recent classifications by
\citetads{2015ApJS..219...19L} and \citetads{2014AJ....147..129S}
given in Fig.~\ref{Fig_coolDA}.

\subsection{New cool DC WDs and candidates}
\label{Sect_coolDC}

Featureless spectra, traditionally defined as showing no line
deeper than 5\% of the continuum, belong to the class of DC WDs
\citepads{1993PASP..105..761W}. Our low-resolution spectra allow 
only for a first selection of DC candidates. With higher S/N
observations weak features can be detected in many WDs initially
classified as DC, but true DC WDs do still exist, especially at 
lower effective temperatures \citepads{1993PASP..105..761W}.
Therefore, we expected to find in particular faint DC candidates  
in our target list. Whereas about two thirds of the 226 WDs in the 25\,pc
sample of \citetads{2016MNRAS.462.2295H} are DA WDs, the DC WDs
represent the second largest fraction with $\approx$17\%.

\begin{figure*}
   \centering
   \includegraphics[width=15.0cm]{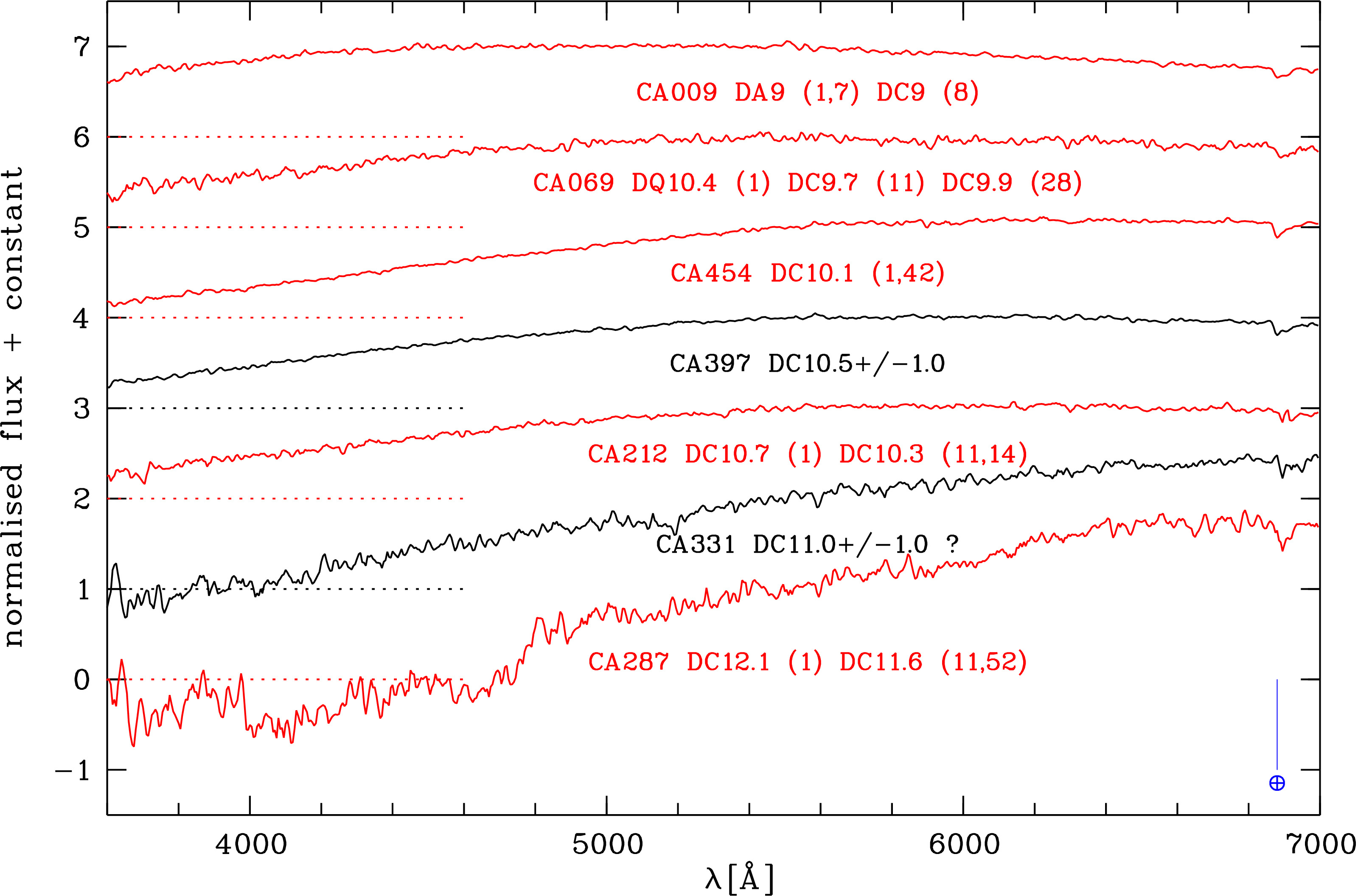}
      \caption{CAFOS spectra of known (red) and new (black) cool DC WDs.
              The spectral types of known objects,
              \object{CA009} = \object{WD 0019+423},
              \object{CA069} = \object{WD 0344+014},
              \object{CA454} = \object{WD 2215+368},
              \object{CA212} = \object{WD 1033+714},
              \object{CA287} = \object{LSPM J1341+0500},
              with typical
              errors of less than about 0.1 subtypes are taken from the
              following references:
              (1) = \citetads{2015ApJS..219...19L},
              (7) = \citetads{1997ApJS..112..527P},
              (8) = \citetads{1984ApJ...276..602G},
              (11) = \citetads{2014AJ....147..129S},
              (14) = \citetads{2008AJ....135.1225H},
              (28) = \citetads{2007AJ....134..252S},
              (42) = \citetads{2012ApJS..199...29G}, and
              (52) = \citetads{2012AJ....143..103S}.
              The spectra are roughly
              sorted by the slope of their continuum.
              A terrestrial absorption band is marked.
              }
         \label{Fig_coolDC}
   \end{figure*}

Among the featureless spectra of known and candidate DC WDs shown
in Fig.\ref{Fig_coolDC}, the coolest known DC WD \object{CA287} 
($\approx$DC12)
and the coolest new candidate \object{CA331} were observed with
only low S/N ratio. Consequently, we consider \object{CA331}
as doubtful 
DC11 candidate (with question mark). Our second new
DC WD, \object{CA397}, shows a very smooth featureless spectrum 
and was classified as DC10.5$\pm$1.0 based on its comparison with
the known DC9-DC11 WDs observed by us. With our assigned uncertainty
of 1.0 spectral subtypes we again took into account the slight
variations in the slopes of the spectra of comparison objects as
well as the range of their spectral types given by different authors.

\subsection{A new DB WD}
\label{Sect_DB}

\begin{figure*}
   \centering
   \includegraphics[width=15.0cm]{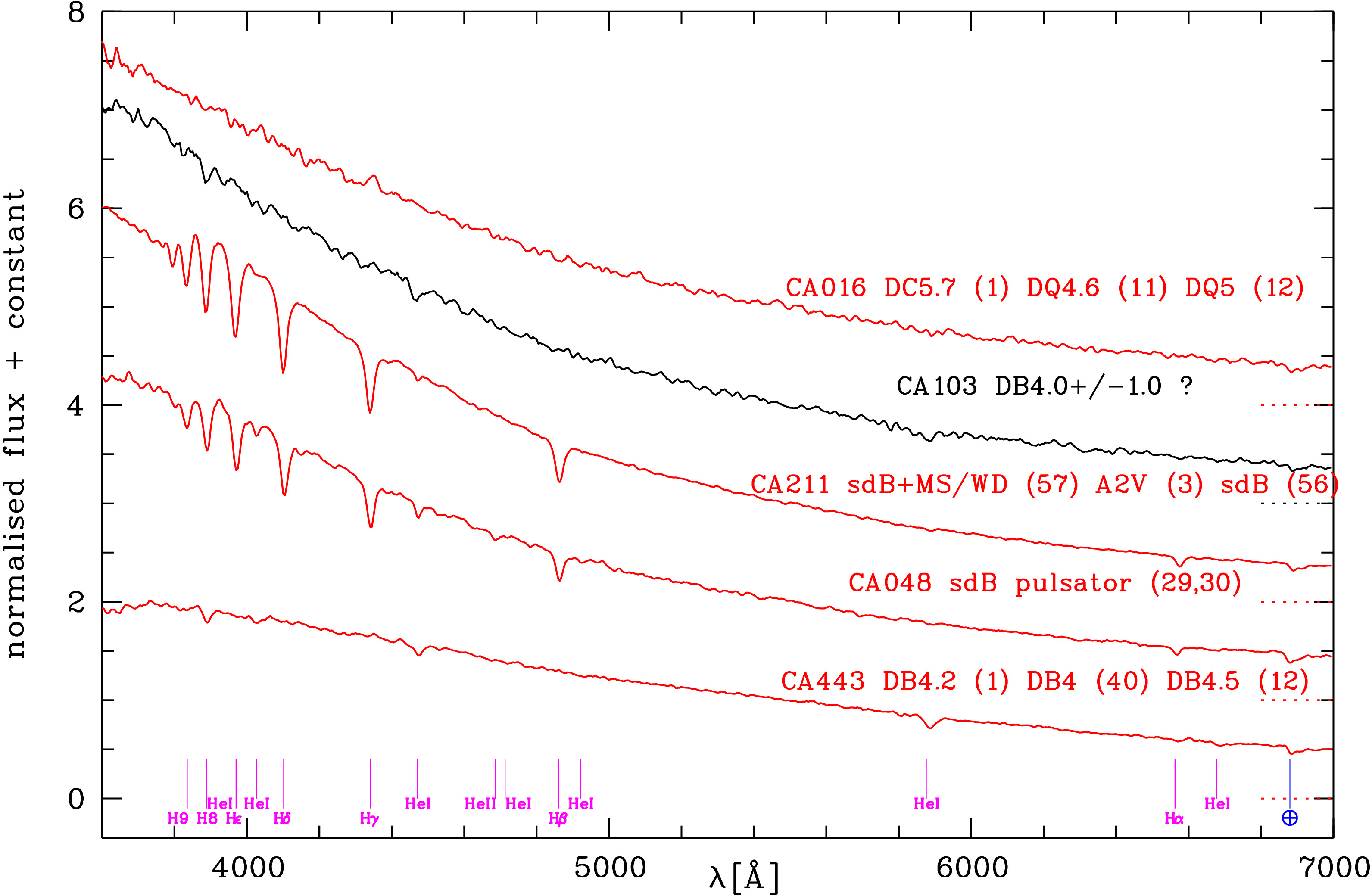}
      \caption{CAFOS spectra of known (red) cool DB and DC WDs and
              sdB stars,
              \object{CA016} = \object{WD 0038+555},
              \object{CA211} = \object{PG 1032+406},
              \object{CA048} = \object{GD 1053},
              \object{CA443} = \object{WD 2147+280},
              and of a new (black) cool DB candidate.
              The spectral types of known objects with typical
              errors of less than about 0.1 subtypes are taken from the
              following references:
              (1) = \citetads{2015ApJS..219...19L},
              (3) = \citetads{2015RAA....15.1095L},
              (11) = \citetads{2014AJ....147..129S},
              (12) = \citetads{2003ApJS..147..145H},
              (29) = \citetads{2014A&A...563A..79R},
              (30) = \citetads{2012MNRAS.421..181R},
              (56) = \citetads{2010A&A...513A...6O}, and
              (57) = \citetads{2015A&A...576A..44K}.
              The spectra are roughly
              sorted by the slope of their continuum. The Balmer lines,
              He\,I and He\,II lines, and a terrestrial absorption band 
              are marked on the bottom.
              }
         \label{Fig_DB}
   \end{figure*}

Those WDs that show only helium lines in their spectra
are called DO (with effective temperatures generally above
45\,000\,K and He\,II lines) or DB (with lower temperatures,
corresponding to the range DB2-DB5, and only He\,I lines)
\citepads{1993PASP..105..761W}. There are only two DB WDs among
the 226 WDs in the 25\,pc sample of \citetads{2016MNRAS.462.2295H}.
Therefore,
we had only one known DB WD (\object{CA443} = \object{WD 2147+280})
selected as a comparison object in
our target list. Nevertheless, we were able to identify one new
DB candidate (\object{CA103}) in our spectroscopic data set. The
spectra of both objects are shown in Fig.~\ref{Fig_DB} together
with the spectra of two known sdB stars and a known DC WD.

Concerning the spectra and marked absorption lines in Fig.~\ref{Fig_DB},
we note that the H8 line of hydrogen and the He\,I line at 3889\,{\AA}
are located very close to each other \citepads{1992A&A...265..781S}. 
While we see the H8 line as part of the dominating Balmer line series 
in the spectra of the two hot subdwarfs shown in Fig.~\ref{Fig_DB}
for comparison, we can see the He\,I line at 3889\,{\AA} 
together with two other He\,I 
lines (at 4471\,{\AA} and 5876\,{\AA}) in the spectra
of both the known $\approx$DB4 WD (\object{CA443} = \object{WD 2147+280})
and of our new
DB3.5 candidate \object{CA103}. A featureless spectrum of a known
$\approx$DC6/DQ5 WD is shown on top to demonstrate
the weakness of the He\,I lines in the spectra of the DB WDs. As 
one can see in Fig.7 of \citetads{1993PASP..105..761W}, the classical 
DB WDs exhibit more He\,I lines at higher temperatures (DB2) than at 
the cool end of the DB sequence (DB5), where only few He\,I lines remain
visible, including those at 3889\,{\AA} and 4471\,{\AA}. This led us 
classify \object{CA103} as a not much earlier type than \object{CA443}
(= \object{WD 2147+280}),
despite of the rather different slopes of their continuum, with an
uncertainty of one spectral subclass and a question mark because of
the weakness of the spectral features with respect to the noise.

\subsection{A new DQ+DAZ WD or G-type carbon dwarf?}
\label{Sect_DQAZ}

The third largest fraction ($\approx$11\%) of WDs in the 25\,pc sample
\citepads{2016MNRAS.462.2295H} is formed by 24 DQ WDs with atomic
or molecular carbon features in their spectra. This fraction is
followed by 10 DZ WDs ($\approx$4\%), which show mainly other 
metallic lines (Ca, Mg, Fe, Na). 
Another 10 WDs in
the 25\,pc sample form a sub-group in the DA WD fraction, the
so-called DAZ WDs with strong Balmer lines and weak metal lines.

   \begin{figure*}
   \centering
   \includegraphics[width=15.0cm]{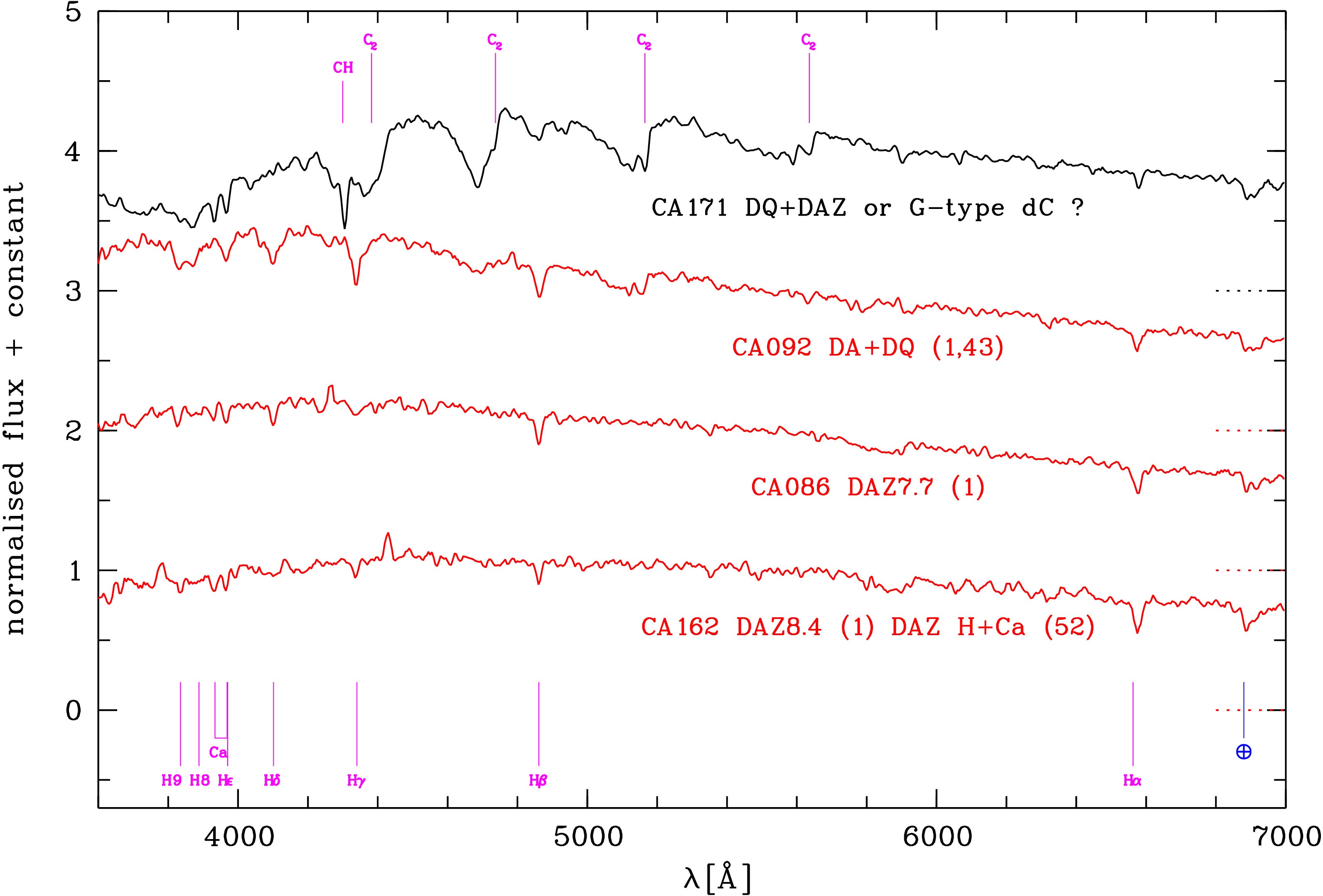}
      \caption{CAFOS spectra of known (red) DAZ and DA+DQ WDs,
              \object{CA092} = \object{GD 73},
              \object{CA086} = \object{LSPM J0543+3637},
              \object{CA162} = \object{WD 0920+012},
              and of a new DQ+DAZ or G-type dC candidate (black).
              The spectral types of known objects with typical
              errors of less than about 0.1 subtypes are taken from the
              following references:
              (1) = \citetads{2015ApJS..219...19L},
              (43) = \citetads{2012ApJ...745L..12V}, and
              (52) = \citetads{2012AJ....143..103S}.
              Locations of the Balmer lines, the Ca H and K lines, and 
              of a terrestrial absorption band are marked on the bottom.
              The bandheads of four C$_2$ Swan bands and the location of
              the CH band are marked on top.
              }
         \label{Fig_DQ_dC}
   \end{figure*}

Among our spectra we found one peculiar WD candidate 
\object{CA171} with
characteristic spectral features of several WD classes.
It resembles partly
that of the known DA+DQ close binary \object{CA092} (= \object{GD 73})
\citepads{2012ApJ...756L...5V} (Fig.~\ref{Fig_DQ_dC}).
The spectrum of \object{CA171} shows as most prominent features
all four Swan bands (at 4382\,{\AA}, 4736\,{\AA}, 5165\,{\AA}, and
5635\,{\AA}) of molecular carbon typically seen in DQ WDs
\citepads{1993PASP..105..761W}
but also the G band of CH at 4300\,{\AA}. In that respect, it looks 
very similar to the peculiar DQP8 WD \object{G 99-37} 
(= \object{WD 0548-001}) in Fig.~19 of \citetads{1993PASP..105..761W}.
However, the spectrum of \object{CA171} shows as additional features
the H$\alpha$ (and H$\beta$) and the Ca H and K lines, typical for
DAZ WDs, two of which are shown for comparison in the lower part
of Fig.~\ref{Fig_DQ_dC}. Therefore, we considered \object{CA171}
as a DQ+DAZ binary WD candidate.

Our alternative classification of this relatively bright 
($G\approx13.35$\,mag) and moderately blue ($G-J\approx+1.27$\,mag,
$J-K_{\rm s}\approx+0.42$\,mag) 
high proper motion 
($\approx166$\,mas/yr) star 
is that of a so-called G-type carbon dwarf (dC) star. This class of
objects was investigated 
on the basis of 
spectroscopic data from the 
SDSS 
by \citetads{2013ApJ...765...12G}.
An example spectrum is presented
in his fig.~2, 
which exhibits the
same main spectral features as observed
by us with lower resolution and S/N for \object{CA171}.

\subsection{Other spectroscopically rejected (candidate) WDs}
\label{Sect_rejected}

\begin{figure*}
   \centering
   \includegraphics[width=15.0cm]{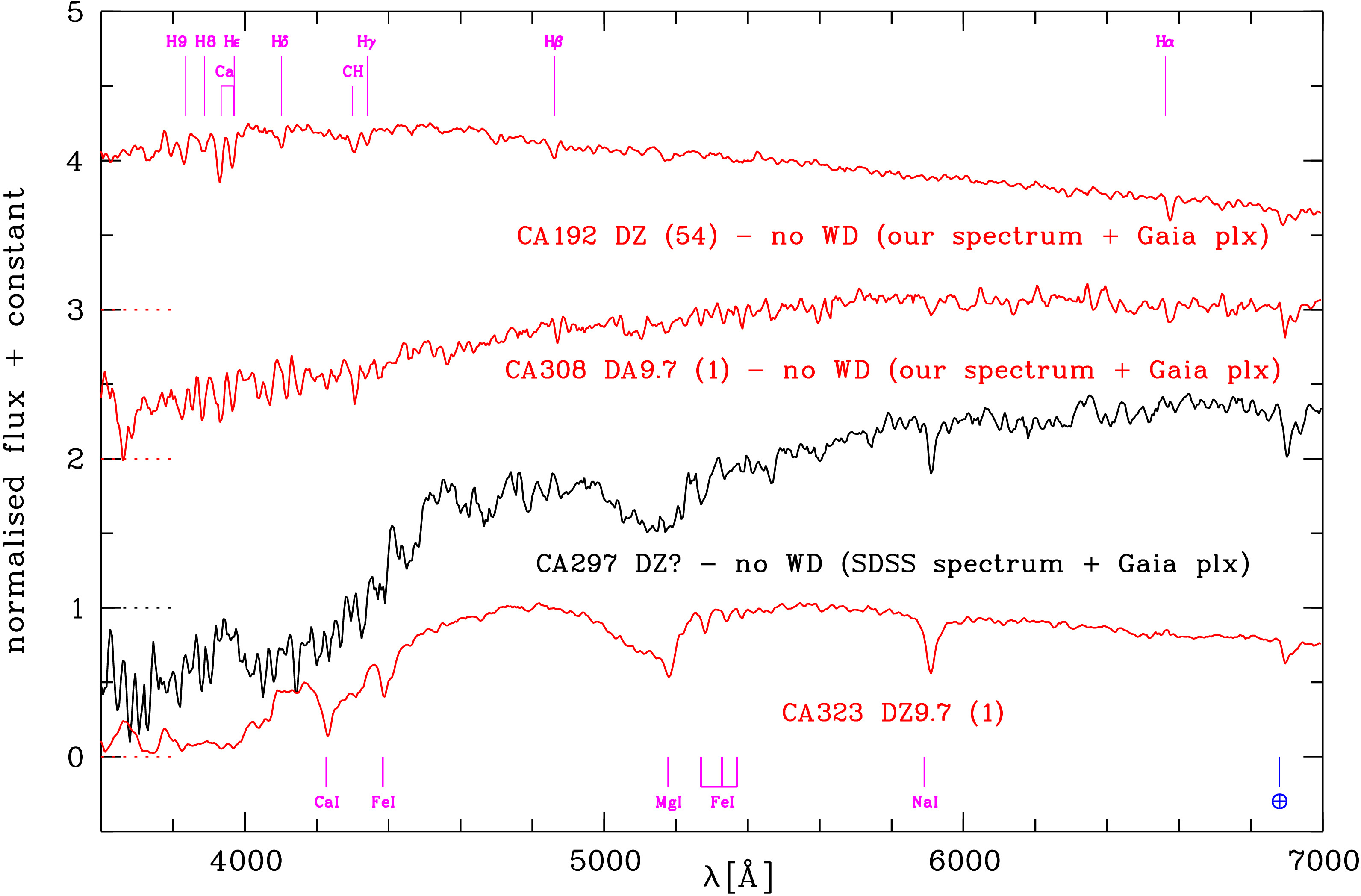}
      \caption{CAFOS spectra of two rejected known WDs (plotted in red
              on top), 
              \object{CA192} = \object{WD 1004+665},
              \object{CA308} = \object{LSPM J1445+2527},
              a rejected new DZ candidate (black), and
              a known DZ WD.
              \object{CA323} = \object{WD 1532+129},
              (plotted in red on bottom).
              The spectral types of known objects with typical
              errors of less than about 0.1 subtypes are taken from the
              following references:
              (1) = \citetads{2015ApJS..219...19L} and
              (54) = \citetads{2012AJ....143..103S}.
              Locations of the strongest metallic lines in the 
              spectrum of the known DZ WD and of a terrestrial 
              absorption band are marked on the bottom.
              Those of the Balmer lines, the Ca H and K lines, and 
              the CH band are marked on top.
              }
         \label{Fig_rejected_WDs}
   \end{figure*}

During our visual classification by 
comparing the spectra 
of
new WD candidates with those of known WDs, we mentioned two 
previously
known WDs, \object{CA192} (= \object{WD 1004+665})
and \object{CA308} (= \object{LSPM J1445+2527}),
that 
failed our own WD classification. In their spectra
shown in the upper part of Fig.~\ref{Fig_rejected_WDs} we identified 
weak Balmer lines
in addition to the Ca H and K lines as well as the 
G band of CH (all lines are marked on top of Fig.~\ref{Fig_rejected_WDs}).
Therefore, we classified these 
objects as
F- or G-type stars.

The spectrum of our faintest ($G\approx17.3$\,mag) 
object \object{CA297} turned out
to be peculiar and relatively red among our targets. We found
the known DZ9.7 WD \object{CA323} 
(= \object{G 137-24} = \object{WD 1532+129})
as its closest match. According to
\citetads{2017MNRAS.467.4970H} this is
the nearest known DZ WD.

The most prominent lines of \object{CA323} as listed 
in \citetads{2004AJ....127.1702K} are marked 
at the bottom 
of Fig.~\ref{Fig_rejected_WDs}. The very asymmetric profile
of the most prominent Mg\,I line at 5170\,{\AA} was 
mentioned in this 
paper. Adopting the explanation 
by \citetads{1980A&A....86..139W} this is 
the result of quasi-static 
van der Waals broadening. This broad feature, as well as some
other characteristic lines (Na\,I and Fe\,I) in the 
spectrum of \object{CA323} seem also
to be present in the 
spectrum of \object{CA297}.

However, 
in \citetads{2015ApJ...813...26F} 
\object{CA297} (= \object{SDSS J140921.10+370542.6}) is
listed as a runaway M dwarf candidate.
Its SDSS spectrum is of higher S/N than our CAFOS spectrum
and clearly rules out a WD. It was classified as that of a 
K7 (SDSS DR14 spectroscopic data base) or
M0V star \citepads[SIMBAD reference:][]{2011AJ....141...97W}.

\section{Absolute magnitudes from Gaia DR2 parallaxes}
\label{Sect_Plxg}

\begin{figure}
   \centering
   \includegraphics[width=\hsize]{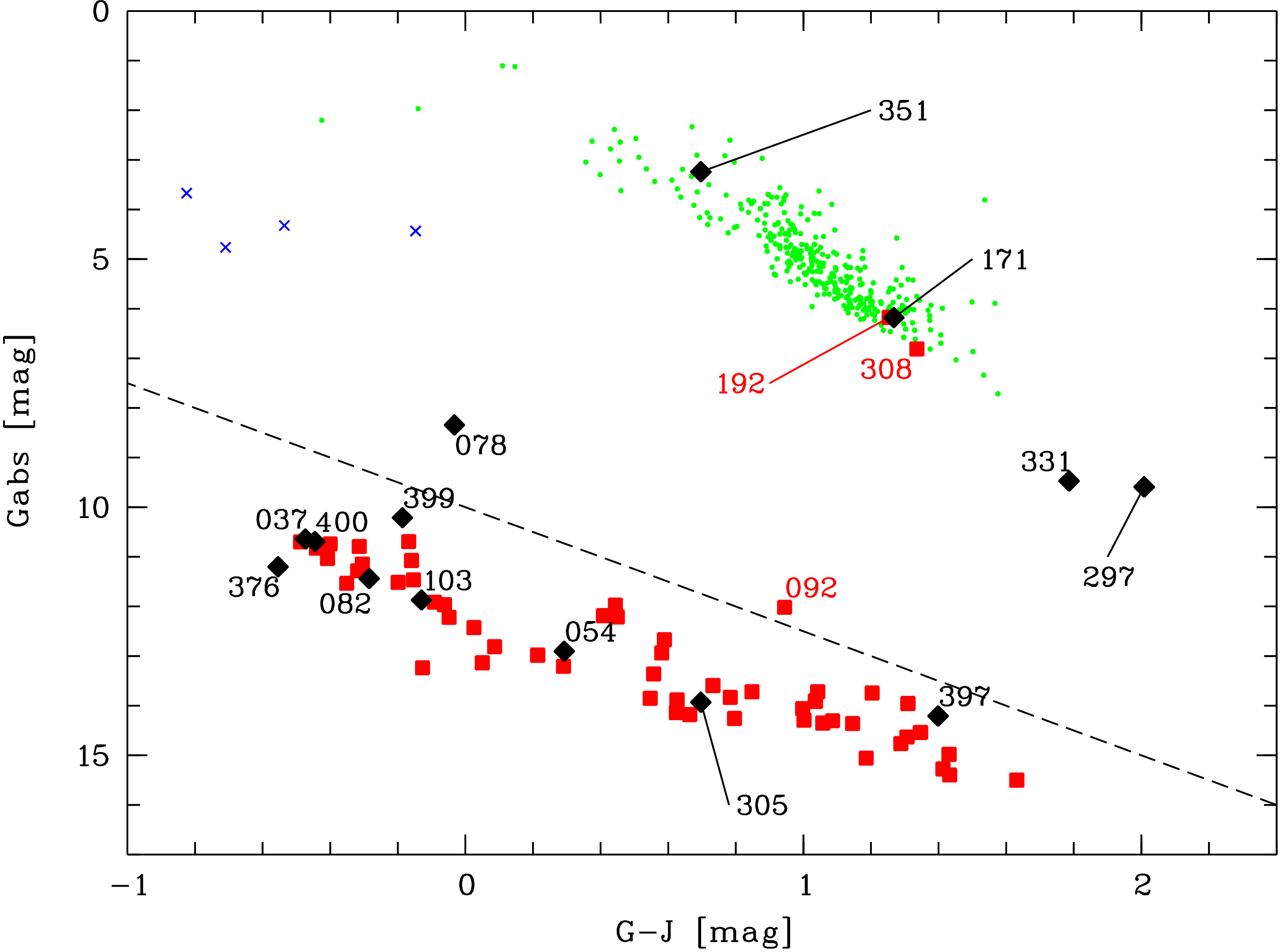}
      \caption{Optical-to-NIR (Gaia DR2 - 2MASS)
               CMD of the objects in our spectroscopic sample:
               red squares = previously known WDs (including two 
               rejected), black lozenges = new WD candidates 
               (confirmed and rejected), blue crosses = hot
               subdwarfs (sdO and sdB), green dots = all
               remaining stars (mainly representing FGK subdwarfs).
               New WDs and candidates, rejected known WDs, and
               one known WD binary are marked by their CA numbers.
               The dashed line corresponds to one of two colour-magnitude 
               limits for WDs used in \citetads{2018A&A...613A..26S}.
              }
      \label{Fig_MG_G_J}
   \end{figure}

\begin{figure}
   \centering
   \includegraphics[width=\hsize]{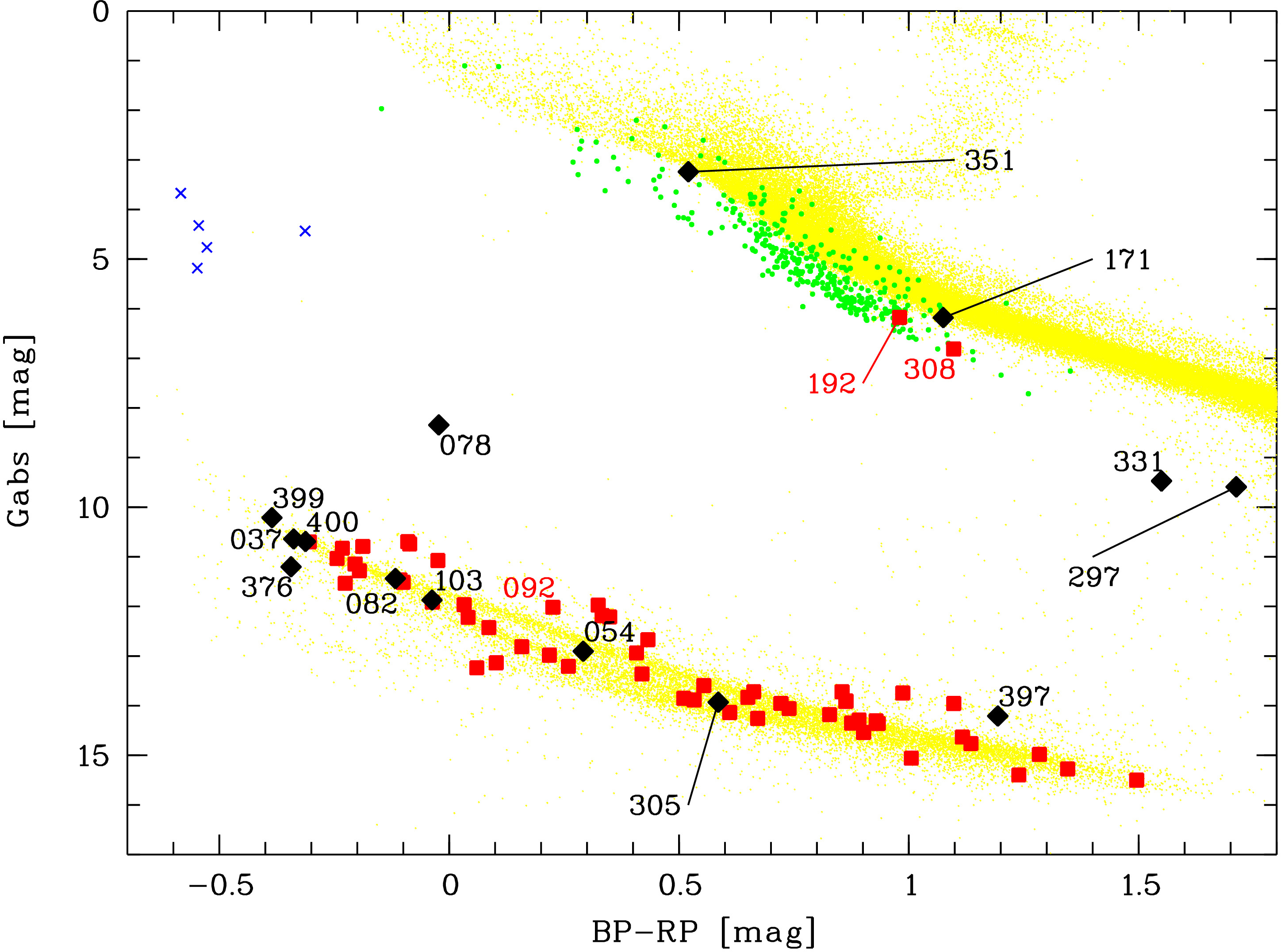}
      \caption{Optical (Gaia DR2) CMD of the 
              same objects and with the same symbols as in 
              Fig.~\ref{Fig_MG_G_J}
              Yellow dots in the background show all
              of the 242,582 objects of the high-quality Gaia 
              DR2 100\,pc sample \citepads{2018arXiv180409366L}
              that fall in this region of the CMD.
              }
      \label{Fig_MG_BP_RP}
   \end{figure}

Almost all of our 410 blue proper motion stars have measured 
parallaxes in Gaia DR2
\citepads{2018arXiv180409365G}. The four stars that are lacking
Gaia DR2 parallaxes were all classified by us (and in the literature) 
as non-WDs. In Figs.~\ref{Fig_MG_G_J} and \ref{Fig_MG_BP_RP} we show
optical-to-NIR and optical CMDs,
where the absolute Gaia magnitudes on the y-axis were computed from 
Gaia DR2 magnitudes and parallaxes. The optical-to-NIR colours shown in
Fig.~\ref{Fig_MG_G_J} were determined from the observed
Gaia DR2 $G$ magnitudes and 2MASS $J$ magnitudes 
of our targets, whereas the x-axis in Fig.~\ref{Fig_MG_BP_RP}
represents their Gaia DR2 $BP-RP$ colours.
We note that the majority of our spectroscopic targets
(shown by green dots) is dominated by subdwarfs
rather than normal dwarf stars (compare with the full Gaia DR2 100\,pc
sample shown in Fig.~\ref{Fig_MG_BP_RP} by yellow dots), because of
our combined proper motion and colour selection (Sect.~\ref{Sect_sample}).

All previously known WDs, except for two objects (\object{CA192}
and \object{CA308}) already found to be non-WDs according to 
our spectroscopic classification (Sect.~\ref{Sect_rejected}),
occupy the lower left part of Fig.~\ref{Fig_MG_G_J} as expected for WDs.
Compared to the WD colour-magnitude limit used in the study
of \citetads{2018A&A...613A..26S} based on parallaxes and
$G$ magnitudes from the Tycho-Gaia Astrometric Solution (TGAS) of
Gaia DR1 (dashed line in Fig.~\ref{Fig_MG_G_J}), 
only one known WD (\object{CA092} with our observed
spectrum shown in Fig.~\ref{Fig_DQ_dC}) lies slightly
above this line. This was not unexpected
since \object{CA092} (= \object{GD 73}) is known as a close WD
binary \citepads{2012ApJ...756L...5V}. However, in the
optical CMD
in Fig.~\ref{Fig_MG_BP_RP} it does neither appear over-luminous 
nor too red compared to other known WDs in our sample.

Concerning our new WDs and WD candidates, 
the weak DA6 candidate \object{CA351} (Sect.~\ref{Sect_coolDA}), 
the doubtful DC11 candidate \object{CA331} (Sect.~\ref{Sect_coolDC}), 
the DQ+DAZ binary WD candidate \object{CA171} (Sect.~\ref{Sect_DQAZ}), 
and our DZ candidate \object{CA297} with an SDSS spectrum that already
ruled out its WD nature (Sect.~\ref{Sect_rejected}) are clearly rejected
as WDs by their Gaia DR2 parallaxes and fall in the 
subdwarf sequences
of Figs.~\ref{Fig_MG_G_J} and \ref{Fig_MG_BP_RP}. 
This is in particular interesting for 
\object{CA171}, which we alternatively classified as a G-type 
carbon dwarf (Sect.~\ref{Sect_DQAZ}). \citetads{2013ApJ...765...12G} 
noted that G-type dC stars
appear bluer than similar mass main sequence stars because of the
absorption bands of molecular carbon. They estimated their
absolute $i$-band
magnitudes as $M_i\approx7.25$\,mag, similar to those of K7-M0 dwarfs.
This is consistent with the $G\approx6.2$\,mag of \object{CA171}.

For all other new WDs and most of the remaining uncertain candidates
including \object{CA054}, the WD candidate with the largest uncertainty 
in our spectral classification (DA5.0$\pm$2.0) and our DB3.5 
candidate \object{CA103}, the Gaia DR2 parallaxes fully confirm
their WD status. These new WDs are located along the WD sequences 
in both Figs.~\ref{Fig_MG_G_J} and \ref{Fig_MG_BP_RP}, with a 
concentration at the blue end, especially in the optical CMD
(Fig.\ref{Fig_MG_BP_RP}).

One new WD candidate, \object{CA078} (with our uncertain spectral
classification of DA3.0$\pm$1.5), deserves special attention. It is
located about 3\,mag above the WD sequence at about zero colour indices
(also typical of A0 stars)
in both Figs.~\ref{Fig_MG_G_J} and \ref{Fig_MG_BP_RP}.
We checked the Gaia DR2 data of this object concerning astrometric
and photometric quality flags as described
in \citetads{2018arXiv180409366L}. We found that \object{CA078}
with a highly significant Gaia DR2 parallax of 13.97$\pm$0.05\,mas
and very large $BP$ and $RP$ mean flux over error ratios ($\approx$459
and $\approx$763) falls well in their ``Selection A'',
which represents a basic 100\,pc sample.
Its astrometric excess noise is small (0.145\,mas). However,
\object{CA078} has a magnitude of $G\approx12.6$\,mag, and the
behaviour of the astrometric excess noise may be less discriminating
at $G\lesssim15$\,mag according to \citetads{2018arXiv180409366L}.
The alternative unit weight error $u$ used by these authors
for their second ``Selection B'' is based on other parameters
characterising the quality of the astrometric solution. It can be
computed for \object{CA078} as $u\approx2.27$, which is still below the
limit of $u\approx5$ at the given $G$
magnitude \citepads[see Fig.C.2 in][]{2018arXiv180409366L}.
Their final and third criterion concerning the flux excess factor
was taken from \citetads{2018arXiv180409378G}
and led to their ``Selection C''. As this last criterion mainly concerns
photometric problems of faint objects in crowded fields it poses no
problems for \object{CA078}.

As a member of the high quality 100\,pc sample corresponding
to ``Selection C'' of
\citetads{2018arXiv180409366L}, the clearly overluminous
WD \object{CA078} can also be seen in their cleaned CMD (their
Fig.~C.1, right panel) as one of very few objects in the region between
the main sequence and the WD sequence. In fact, it appears there as one of
only two objects with colour indices $BP-RP\approx0.0$\,mag
and 3-4\,mag above
the WD sequence, but also 3-4\,mag below a small group of five possible
A-type subdwarfs located about 2\,mag below the main sequence,
in their cleaned CMD
(yellow dots in our Fig.~\ref{Fig_MG_BP_RP}).
This is a remarkable uncommon object found among our ten
spectroscopically confirmed new WDs in a small sample of
410 blue proper motions stars and the much larger ``Selection C''
sample of \citetads{2018arXiv180409366L} consisting of 242,582 objects.
We further discuss the possible binary status of \object{CA078}
in Sect.~\ref{Sect_CA078}.

\section{WD classification from Gaia DR2 colours}
\label{Sect_BP_RP}

\begin{figure}
   \centering
   \includegraphics[width=\hsize]{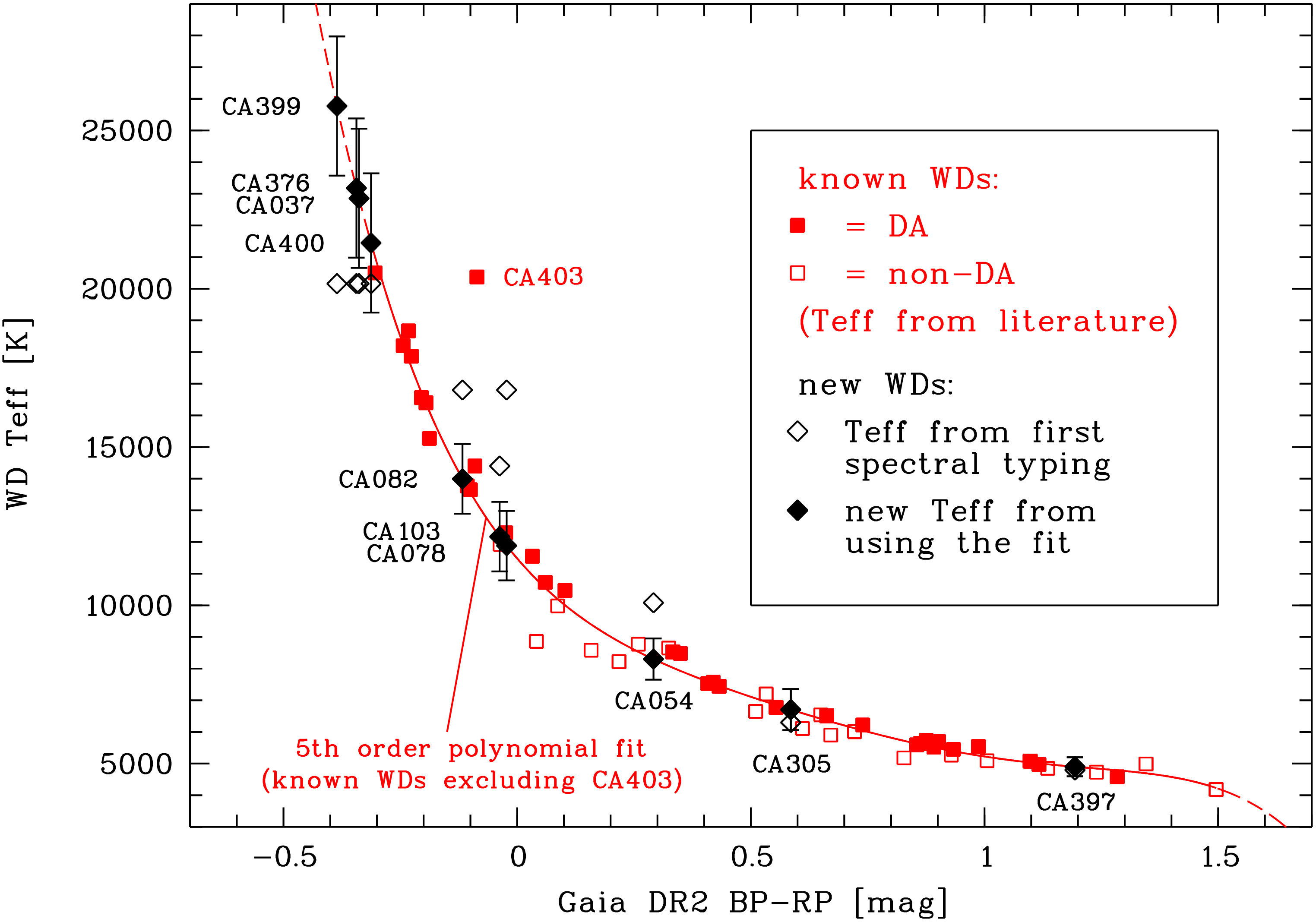}
      \caption{WD effective temperatures (from literature)
               as a function of Gaia colours $BP-RP$ for
               known (and confirmed) DA (filled red squares)
               and non-DA (open red squares) WDs. The solid 
               and dashed red lines show the 5th order polynomial fit
               for these objects excluding \object{CA403}
               (= \object{WD 1943+163})
               (see text) and its extrapolations.
               Black open lozenges mark the effective temperatures
               of new (and confirmed) WDs corresponding to their
               initial spectroscopic classification 
               (Sect.~\ref{Sect_class}) 
               as a function of their Gaia colours.
               Black filled lozenges mark their newly derived
               effective temperatures using the fit with known
               WDs (error bars are explained in the text).
               The fitted values of new WDs are marked by 
               their CA names.
              }
      \label{Fig_Teff_BP_RP}
   \end{figure}

To further improve the classification of our new WDs, we investigated
the effective temperatures available from the literature of all known 
(and confirmed by Gaia DR2 parallaxes) WDs in our sample
as a function of the measured Gaia DR2 colours $BP-RP$ and found a strong 
correlation between these two quantities (red filled and open squares
in Fig.~\ref{Fig_Teff_BP_RP}).
The effective temperatures were mainly taken from
\citetads{2015ApJS..219...19L}, with additional data from
\citetads{2011ApJ...743..138G},
\citetads{2014AJ....147..129S}
\citetads{2012MNRAS.425.1394K}, and
\citetads{2015RAA....15.1095L}. Note that the binary WD \object{CA092}
was not included here, as its effective temperature was lacking.
The red solid line shows a 5th order polynomial fitting these data of
the previously known WDs.
Excluded from the fit was the WD
\object{CA403} (= \object{WD 1943+163}), which is
a clear outlier.
Among all known WDs in our sample it is 
the only object with a Gaia duplicated source
flag (=1). According to Gaia DR2
documentation, 
such a 
flag may indicate
``observational, cross-matching or processing problems, or stellar
multiplicity, and probable astrometric or photometric problems in
all cases''.

The filled black lozenges show the newly determined effective temperatures
of our new WDs 
applying 
this polynomial fit, where we
had to extrapolate 
at the hot end (dashed line). Their initially estimated rough effective
temperatures according to our former spectroscopic classification are
shown by open black lozenges. The maximum deviations of the known WDs 
(red filled and open squares)
from the fitted curve 
amount to 
$\pm$1100\,K in the colour
interval ($-0.3 \lesssim BP-RP \lesssim 0.0$)\,mag, 
where the curve is steeper, but only $\pm$650\,K in the 
colour interval ($+0.2 \lesssim BP-RP \lesssim +0.9$)\,mag
and $\pm$300\,K in the    
colour interval ($+0.9 \lesssim BP-RP \lesssim +1.5$)\,mag,
where the curve is more and more flattening. We conservatively estimated
the uncertainties of our newly determined effective temperatures for
new WDs to be as large as these maximum deviations in the corresponding
colour intervals, but applied an additional factor of 2.0 for
the very blue ($-0.4 \lesssim BP-RP \lesssim -0.3$)\,mag colours, 
where we had to extrapolate.

The resulting uncertainties in effective temperatures 
yield also
smaller errors in the 
determination of the spectral type from
$\pm$0.2 subtypes
for our new hot WDs to about $\pm$0.8 subtypes for our new cool WDs.
Our initial spectral types
and our finally adopted spectral types of higher precision agree within
their assigned uncertainties
(Table~\ref{Tab_10WDs}).

%
%
\begin{table*}
\caption{Gaia DR2 astrometry, photometry, and tangential velocities, initial spectral types from CAFOS spectroscopy, and estimated effective temperatures and our corresponding adopted spectral types of ten new WDs from Gaia colours.} 
\label{Tab_10WDs}
\centering
\fontsize{6.4pt}{0.90\baselineskip}\selectfont
\begin{tabular}{@{}llrrrrrc@{}rccc@{}r@{}}     
\hline\hline
CA  & Alternative & RA\tablefootmark{a} & DE\tablefootmark{a} & $Plx$\tablefootmark{a} &  $pm$RA\tablefootmark{a} & $pm$DE\tablefootmark{a} & $G$\tablefootmark{a} &  $BP-RP$\tablefootmark{a} & initial & $T_{\rm eff}$ &adopted  & $v_{\rm tan}$\tablefootmark{b} \\
No. & Names &  &  &  &  &  &  &  &  SpT &   & SpT \\
    &  & [degrees] & [degrees] & [mas] & [mas/yr] & [mas/yr] & [mag] & [mag] &  & [K] &  & [km/s] \\
\hline
037 & \object{GD 277} &  22.350203 & +51.145903 &  26.19$\pm$0.05 &  +32.96$\pm$0.09 & -110.58$\pm$0.09 &13.550 &-0.338 & DA2.5$\pm$1.0 & 22857$\pm$2200 & DA2.2$\pm$0.2 &  21 \\
054 & \object{GD 28} &  37.170803 & +36.760490 &  24.41$\pm$0.06 &  +10.41$\pm$0.10 & -171.67$\pm$0.08 &15.965 &+0.292 & DA5.0$\pm$2.0?&  8301$\pm$650  & DA6.1$\pm$0.5 &  33 \\
078 & \object{2MASS J05005185-0930549} &  75.215835 &  -9.515829 &  13.97$\pm$0.05 &  -48.55$\pm$0.07 & -121.87$\pm$0.07 &12.616 &-0.023 & DA3.0$\pm$1.5?& 11884$\pm$1100 & DA4.2$\pm$0.5 &  45 \\
082 & \object{2MASS J05280449+4105253} &  82.019148 & +41.090348 &  18.31$\pm$0.08 &  +66.30$\pm$0.10 &   -3.13$\pm$0.08 &15.124 &-0.117 & DA3.0$\pm$1.0 & 13994$\pm$1100 & DA3.6$\pm$0.3 &  17 \\
103 & \object{2MASS J07035743+2534184} & 105.988994 & +25.571520 &  39.06$\pm$0.03 &  -58.29$\pm$0.06 &  -55.69$\pm$0.05 &13.912 &-0.037 & DB3.5$\pm$1.0?& 12168$\pm$1100 & DB4.1$\pm$0.4 &  10 \\
305 & \object{LP 740-47} & 220.824701 & -14.621474 &  34.89$\pm$0.08 & -225.37$\pm$0.14 & -128.43$\pm$0.12 &16.218 &+0.585 & DA8.0$\pm$1.0 &  6704$\pm$650  & DA7.5$\pm$0.8 &  35 \\
376 & \object{HD 166435 B} & 272.331404 & +29.956111 &  40.90$\pm$0.04 &  +66.93$\pm$0.06 &  +74.25$\pm$0.07 &13.143 &-0.344 & DA2.5$\pm$1.0 & 23180$\pm$2200 & DA2.2$\pm$0.2 &  12 \\
397 & \object{LSPM J1919+4527} & 289.900277 & +45.463443 &  35.59$\pm$0.04 & -108.92$\pm$0.08 & +312.59$\pm$0.09 &16.452 &+1.194 &DC10.5$\pm$1.0 &  4901$\pm$300  &DC10.3$\pm$0.7 &  44 \\
399 & \object{GD 221} & 291.858997 & +10.118643 &  12.53$\pm$0.05 &  -80.04$\pm$0.07 & -188.03$\pm$0.06 &14.724 &-0.386 & DA2.5$\pm$1.0 & 25771$\pm$2200 & DA2.0$\pm$0.2 &  77 \\
400 & \object{2MASS J19293865+1117523} & 292.411127 & +11.298312 &  27.34$\pm$0.05 &  +19.15$\pm$0.07 &  +97.83$\pm$0.06 &13.508 &-0.313 & DA2.5$\pm$1.0 & 21445$\pm$2200 & DA2.4$\pm$0.3 &  17 \\
\hline
\end{tabular}
\tablefoot{\fontsize{6.4pt}{0.90\baselineskip}\selectfont
Gaia DR2 coordinates are for (J2000, epoch 2015.5) and were
rounded to 0.000001 degrees, parallaxes and their errors were rounded to
0.01\,mas, proper motions and their errors to 0.01\,mas/yr, magnitudes
and colours were rounded to 0.001\,mag. The initial spectral types
with question marks were uncertain because of low S/N
or possible problems in the flux calibration of the CAFOS spectra.
Further notes on the data:
\tablefoottext{a}{Gaia DR2,}
\tablefoottext{b}{derived from Gaia DR2 parallaxes and proper motions,}
}
\end{table*}

%
%
\begin{table*}
\caption{Gaia DR2 astrometry, photometry, and tangential velocities, initial (from CAFOS) and adopted spectral types of six rejected WDs.}
\label{Tab_6nonWDs}
\centering
\fontsize{6.4pt}{0.90\baselineskip}\selectfont
\begin{tabular}{@{}llrrrrrc@{}rcccr@{}}     
\hline\hline
CA  & Alternative & RA\tablefootmark{a} & DE\tablefootmark{a} & $Plx$\tablefootmark{a} &  $pm$RA\tablefootmark{a} & $pm$DE\tablefootmark{a} & $G$\tablefootmark{a} &  $BP-RP$\tablefootmark{a} & initial  &adopted & Ref & $v_{\rm tan}$\tablefootmark{b} \\
No. & Names &  &  &  &  &  &  &  & SpT & SpT \\
    &  & [degrees] & [degrees] & [mas] & [mas/yr] & [mas/yr] & [mag] & [mag] &  &  &  & [km/s] \\
\hline
171 & \object{LSPM J0937+2803} & 144.359230 &+28.056399  &  3.68$\pm$0.03 &  +63.38$\pm$0.04 & -153.13$\pm$0.04 &13.354 &+1.075 &    DQ+DAZ? &      G-type dC  & 1 &  214 \\
192 & \object{WD 1004+665} & 152.005966 &+66.322420  &  2.19$\pm$0.02 & -153.10$\pm$0.03 & -127.69$\pm$0.04 &14.475 &+0.980 &    no WD   &     $\approx$G  & 1 & 432 \\
297 & \object{SDSS J140921.10+370542.6} & 212.337320 &+37.094950  &  2.89$\pm$0.08 & -154.80$\pm$0.07 &  -73.75$\pm$0.09 &17.290 &+1.712 &     DZ?    &      M0V (SDSS) & 2 & 282 \\
308 & \object{LSPM J1445+2527} & 221.414204 &+25.453752  &  1.44$\pm$0.05 &  -37.82$\pm$0.10 & -155.26$\pm$0.10 &16.012 &+1.097 &    no WD   &     $\approx$G  & 1 & 524 \\
331 & \object{LSPM J1554+1939} & 238.570339 &+19.663440  &  3.47$\pm$0.07 &  +13.26$\pm$0.09 & -179.27$\pm$0.08 &16.770 &+1.550 &DC11.0$\pm$1.0? & $\approx$K  & 1 & 246 \\
351 & \object{2MASS J16345081+2610073} & 248.711612 &+26.168217  &  1.34$\pm$0.04 &  -30.34$\pm$0.05 & -118.39$\pm$0.07 &12.600 &+0.520 & DA6.0$\pm$1.0? & $\approx$F  & 1 & 431 \\
\hline
\end{tabular}
\tablefoot{\fontsize{6.4pt}{0.90\baselineskip}\selectfont
Gaia DR2 coordinates are for (J2000, epoch 2015.5) and were
rounded to 0.000001 degrees, parallaxes and their errors were rounded to
0.01\,mas, proper motions and their errors to 0.01\,mas/yr, magnitudes
and colours were rounded to 0.001\,mag. The initial spectral types
with question marks were uncertain because of low S/N
or possible problems in the flux calibration of the CAFOS spectra.
Further notes on the data:
\tablefoottext{a}{Gaia DR2,}
\tablefoottext{b}{derived from Gaia DR2 parallaxes and proper motions,}
}
\tablebib{\fontsize{7pt}{0.90\baselineskip}\selectfont
(1) this paper (adopted spectral types based on CAFOS
spectra and the location of the objects along the main sequence in
the CMDs (Figs.~\ref{Fig_MG_G_J} and \ref{Fig_MG_BP_RP}),
(2) \citetads{2011AJ....141...97W}.
}
\end{table*}

\section{Discussion}
\label{Sect_discuss}

\subsection{Classification and kinematics of rejected WDs}
\label{Sect_notes_rejected}

Three objects (\object{CA171}, \object{CA308}, and \object{CA351})
out of six rejected WD candidates and known WDs
described in Sects.~\ref{Sect_DQAZ}  and \ref{Sect_rejected} 
were photometrically
classified as FGK stars by \citetads{2010PASP..122.1437P},
and one object 
(\object{CA192}) was listed as K star candidate
in the catalogue of \citetads{2001KFNT...17..409K}. The
WD classification of \object{CA297} was rejected on the basis of
an SDSS M0V spectrum
already noted in Sect.~\ref{Sect_rejected}.

The largest 
tangential velocity among our new WDs
is 77\,km/s 
for \object{CA399}, possibly indicating its membership
in the Galactic thick disk population.
On the other hand, 
all rejected WDs exhibit
tangential velocities between $\approx210$\,km/s and $\approx520$\,km/s,
relating them to the Galactic halo. They are all located at high
Galactic latitudes (between $+40\degr$ and $+71\degr$), whereas the
ten new WDs lie 
near 
the Galactic plane between $-29\degr$ and $+40\degr$.

The Gaia parallax of \object{CA171} (Table~\ref{Tab_6nonWDs}) 
confirms 
its classification as 
a G-type carbon dwarf. With the corresponding distance of $\approx272$\,pc
it is 
a relatively nearby
representative of this class of objects.
The 
vast majority of
G-type dC stars presented by \citetads{2013ApJ...765...12G} are
much fainter than \object{CA171} and show typical total proper
motions below 30\,mas/yr (see his fig.~11). 
Compared to other 
nearby carbon dwarfs \citepads{2018AJ....155..252H},
distance and tangential velocity of \object{CA171}
are typical for the 
Galactic halo population.
But, compared to other carbon dwarfs
it appears much bluer ($J-K_{\rm s}\approx+0.4$\,mag) and 
relatively bright ($J\approx12.1$\,mag) not only in the NIR (2MASS)
but also in the optical ($B=14.59$\,mag, $V=13.55$\,mag) 
\citepads[APASS;][]{2016yCat.2336....0H}.

\subsection{Previous classification of new WDs}
\label{Sect_prev_class}

Three of the new WDs, 
\object{CA037} (= \object{GD 277}),
\object{CA054} (= \object{GD 28}), and
\object{CA399} (= \object{GD 221}),
were already considered as WD candidates 
by \citetads{1980LowOB...8..157G}.
Later on two of them were
classified as non-WDs in \citetads{2010PASP..122.1437P}
based on multi-colour photometry (\object{CA037} as
A7III candidate and \object{CA054} as G8V candidate).
Another new WD, \object{CA397}, was previously
listed as a \textit{Kepler} target star with 
estimated 1.5 solar masses and about solar metallicity
\citepads{2014ApJS..211....2H}. Two of the new WDs, \object{CA054}
and \object{CA305} were listed in 
Luyten's WD catalogue \citepads{1999yCat.3070....0L}, both
with a spectral or colour class ``f''. 

\subsection{URAT parallaxes of new WDs}
\label{Sect_UPC}

For four of the new WDs 
(\object{CA037}, \object{CA103}, \object{CA376}, and \object{CA397}) 
first trigonometric parallaxes of lower precision were already
published before Gaia DR2 in the
URAT \citepads{2015AJ....150..101Z} 
Parallax Catalog \citepads[UPC;][]{2016yCat.1333....0F,2018yCat.1344....0F}
described by \citetads{2016arXiv160406739F,2018AJ....155..176F}.
For two of them, \object{CA037} and \object{CA103}, the UPC parallaxes
are in excellent agreement with the Gaia DR2 values, and for \object{CA397}
the UPC parallax agrees within the typical UPC error bars ($\pm$3-6\,mas)
with the Gaia DR2 parallax. 
Only for
\object{CA376}, the UPC
parallax is 
about 10\,mas 
larger than the Gaia DR2 parallax.

\subsection{Wide binary companions of our new WDs}
\label{Sect_wide_binaries}

Wide binaries were traditionally found 
among 
common proper motion stars
in high proper motion catalogues.
Now, in 
the Gaia era, we can use very accurate measurements of both
proper motions and parallaxes for an almost complete sample of
stars in the solar neighbourhood. 
Therefore, we 
checked our ten new WDs for
stars with common proper motion and parallax in Gaia DR2 data. 
We used a search radius of 3600\,arcsec and selected only
Gaia DR2 sources with very similar parallaxes and then compared
their proper motions. 

In addition to the already known G-type primary of \object{CA376}
\citepads[= \object{HD 166435 B}; see][]{2018A&A...613A..26S} with
an angular separation of about 29.2\,arcsec, we found 
\object{2MASS J19293859+1118050}
a red ($BP-RP\approx+2.62$\,mag) companion of \object{CA400} separated by
only about 12.7\,arcsec.
Its parallax (27.38$\pm$0.07\,mas) is
in perfect agreement with that of \object{CA400}, whereas its
proper motion of (+18.76$\pm$0.11, 99.80$\pm$0.09)\,mas/yr is
only slightly deviating (Table~\ref{Tab_10WDs}), as expected due to
orbital motion \citepads{2018A&A...613A..26S}. We derived an absolute
$J$ magnitude of $\approx$7.3\,mag for this red companion from its 
2MASS photometry and Gaia DR2 parallax corresponding to a spectral type 
of $\approx$M3 according to the relationship between absolute magnitudes 
and spectral types \citepads{2005A&A...442..211S}.

\subsection{One of the nearest DB WDs: CA103}
\label{Sect_CA103}

With a distance of only $\approx$26.60\,pc, \object{CA103} is the second
nearest of our new WDs after \object{CA376} at $\approx$24.45\,pc
(based on Gaia DR2 parallaxes).  There were only two DB WDs in the 25\,pc 
WD sample \citepads{2016MNRAS.462.2295H},
and one of these two, \object{WD 2058+342}, 
is now found to be
at a distance of $\approx$52.51\,pc according to its Gaia DR2 parallax.\footnote{Investigating the Gaia DR2 20\,pc sample, 
\citetads{2018arXiv180512590H} idendified 128 known and 
11 new WDs, whereas 57 (!) former members of the 20\,pc sample 
were found to be located at larger distances.}
Therefore, \object{CA103} may in fact be the second nearest of all DB WDs
after \object{WD 1917-074} at $\approx$10.50\,pc.
 
Among our new WDs (Table~\ref{Tab_10WDs}), \object{CA103} has a very 
small proper 
motion for its distance, and consequently the smallest tangential 
velocity of $\approx$10\,km/s, indicating a young age.
Interestingly, \object{CA103} is listed in the
K2 Ecliptic Plane Input Catalog \citepads[EPIC;][]{2017yCat.4034....0H}.
It represents a relatively bright 
($G\approx13.9$\,mag, $J\approx14.0$\,mag) 
target for various follow-up observations.

\subsection{CA078, a nearby extremely low mass WD candidate}
\label{Sect_CA078}

The multiplicity of nearby WDs, in particular double WDs (DWDs)
and WD main-sequence binaries (WDMS), was investigated by
\citetads{2017A&A...602A..16T}. They listed two unresolved DWDs and
eight unresolved DWD candidates within 25\,pc (mostly DA types
between DA5.6 and DA9.9) in their Table 1.
In addition, one of their only four resolved DWDs (DC10) has a small
angular separation of 1.4\,arcsec. They also compared the observed
multiplicity of WDs with star formation and evolution model predictions
and found a discrepancy for resolved DWDs, 
more than ten of these 
are apparently missing within 20\,pc.

   \begin{figure}
   \centering
   \includegraphics[width=7.8cm]{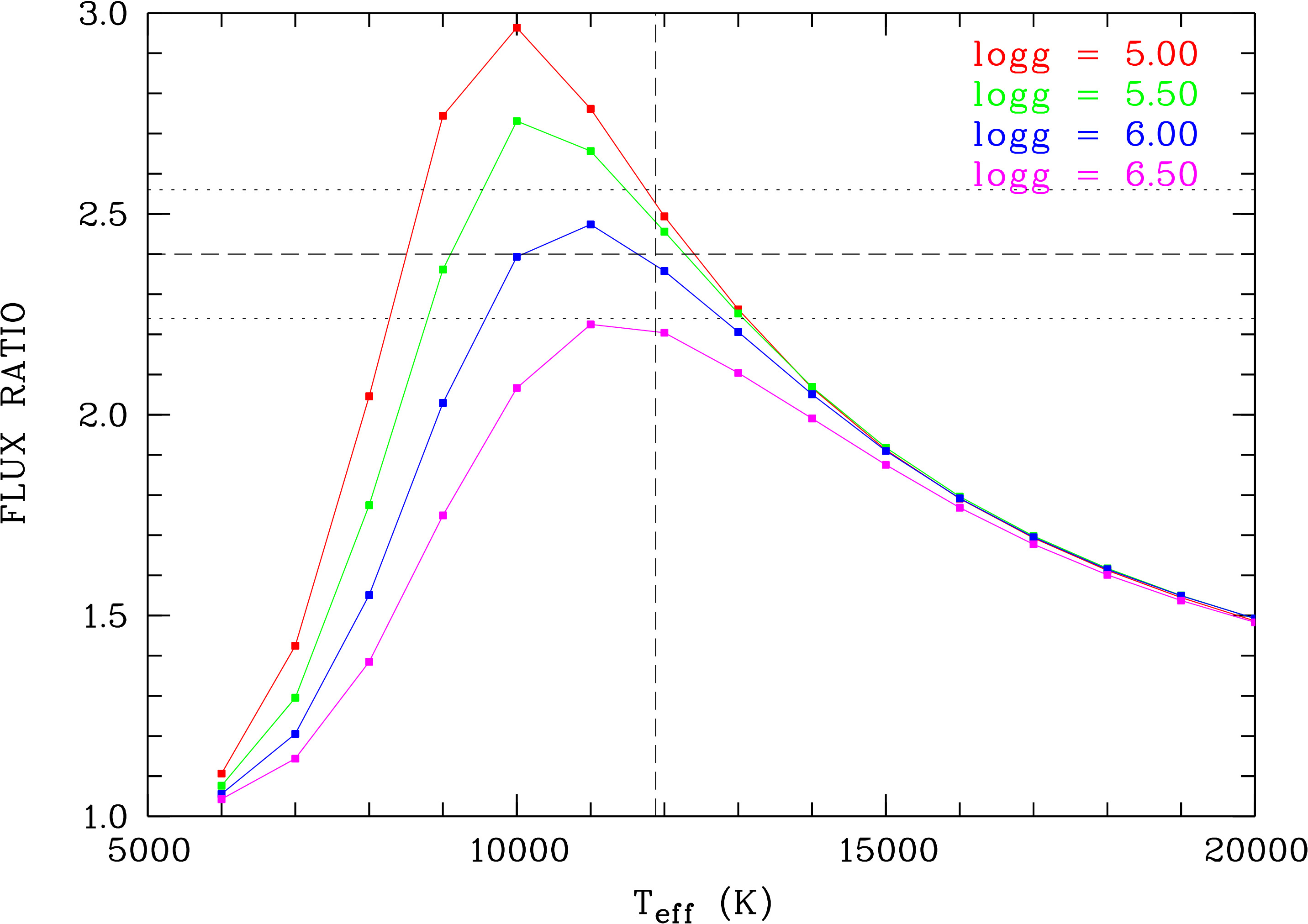}
      \caption{Balmer jump as a function of
               effective temperature measured in low-gravity
               model spectra (small filled squares
               connected by solid lines) between $\log{g}=5.0$ (top)
               and $\log{g}=6.5$ (bottom). The dashed vertical line
               indicates the effective temperature of \object{CA078}
               (Sect.~\ref{Sect_BP_RP}). Dashed and dotted horizontal
               lines indicate our observed value for \object{CA078}
               and our estimated measurement uncertainty.
              }
      \label{Fig_Bjump}
   \end{figure}

Our overluminous WD \object{CA078} lies at a much larger distance
of about 71.6\,pc, where the chances to resolve a possible multiple 
system are lower, and has a slightly earlier spectral type
(DA4.2$\pm$0.5) than the above mentioned unresolved DWDs. 
In addition, an
equal-mass WD binary would appear only 0.75\,mag
brighter than a single WD.
As 
already mentioned 
in Sect.~\ref{Sect_Plxg}, \object{CA078} belongs to the
high-quality Gaia DR2 100\,pc sample and its location about 3\,mag
above the WD sequence in the optical-to-NIR CMD (Fig.~\ref{Fig_MG_G_J})
and optical CMD (Fig.~\ref{Fig_MG_BP_RP}) requires an 
alternative explanation 
different from 
simple binarity.

An elusive class of WDs with low surface gravities of
$5\lesssim \log{g} \lesssim7$ and effective temperatures in the range 
of 8000\,K$\lesssim T_{\rm eff} \lesssim$22\,000\,K are the so-called
extremely low mass (ELM) WDs \citepads{2016ApJ...818..155B}.
With our estimated effective temperature of about 11\,800\,K
well within this interval, can 
the overluminosity of \object{CA078}
be explained with its possible ELM status?
Our Balmer line equivalent width measurements
(Figs.~\ref{Fig_HaHbEW} and \ref{Fig_HbHgEW}), 
in particular of the well-measured
H$\beta$ and H$\gamma$ lines, already provided a hint on the
possible low gravity ($\log{g}\lesssim7$) of \object{CA078}
(Sect.~\ref{Sect_modelspec}).
All known DA3.6-DA4.5 shown
(in Figs.~\ref{Fig_HaHbEW} and \ref{Fig_HbHgEW} 
have much larger H$\beta$ (and H$\alpha$) line widths than \object{CA078},
whereas the H$\gamma$ line widths are similar.

In addition to the 
Balmer line widths, we can also measure the gravity-dependent
Balmer jump. 
We defined it as the flux ratio between the mean spectral energy
densities in the wavelength intervals 4200-4240\,{\AA} and
3700-3740\,{\AA},
which were covered
both by the model and the observed spectra. Those flux ratios were also
measured in spectra normalised to a continuum (instead of measuring in
the original spectra) to avoid a colour term due to calibration
uncertainties of the
instrumental response function for our observed spectra.
In Fig.~\ref{Fig_Bjump} we compare the
size of the Balmer jump in the observed and the
model spectra. The graph indicates a low value of about $\log{g}=6$
of the surface gravity of \object{CA078} with an upper
limit of about $\log{g}=6.5$. This is consistent with an
ELM status of \object{CA078}.

   \begin{figure}
   \centering
   \includegraphics[width=\hsize]{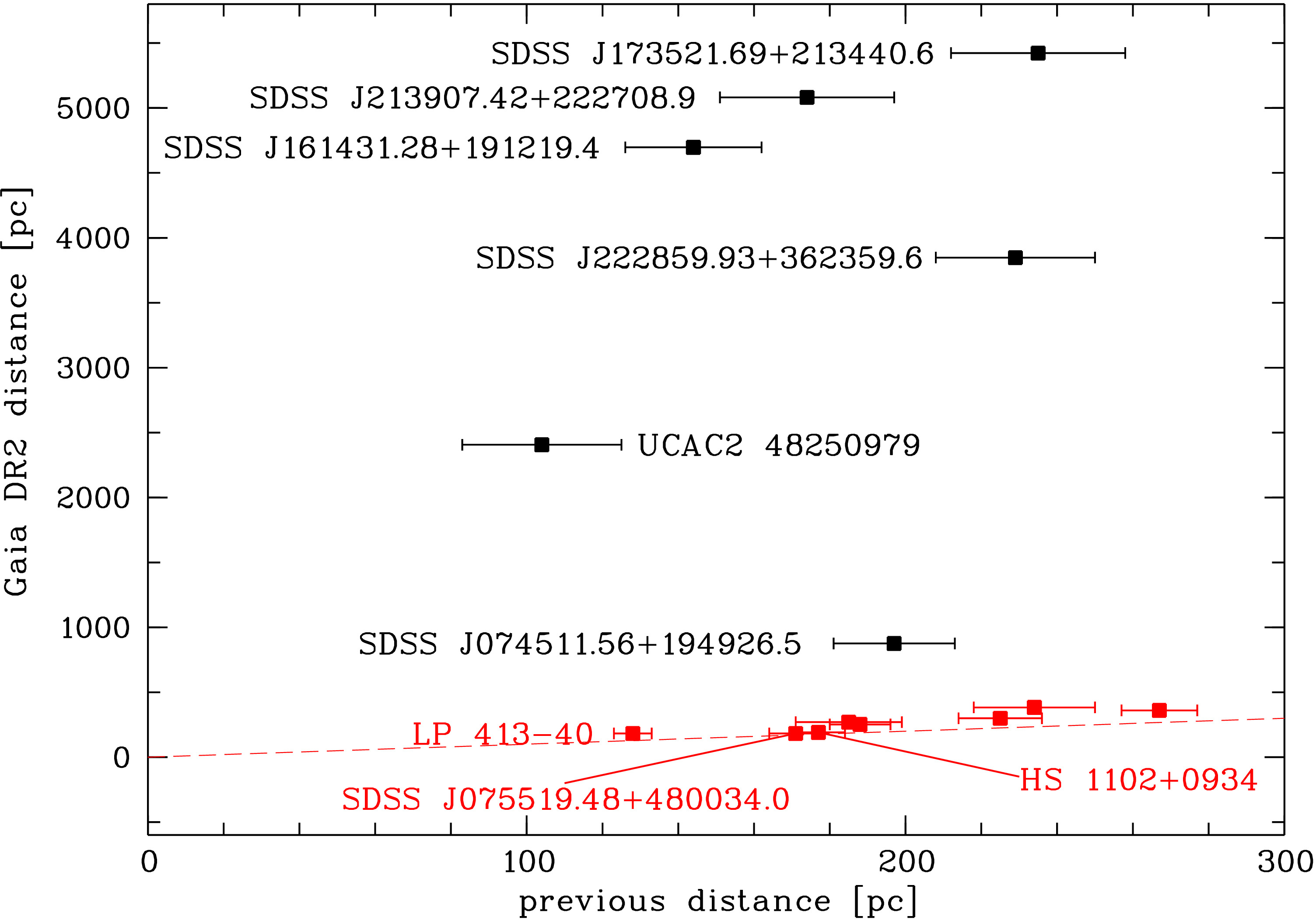}
      \caption{Previous distances of the known nearest ELM WDs
              (see text) and their new distances derived from Gaia DR2
              parallaxes. The red dashed line represents equality.
              Objects shifted by Gaia to much larger distances are
              labelled and shown in black. Other objects 
              are shown
              in red, the three nearest of which are labelled.}
      \label{Fig_ELM}
   \end{figure}

Gaia DR2 and SDSS data were already combined by
\citetads{2018arXiv180504070P} in a study of so-called
sdA stars, including (pre-)ELM WDs of low gravity and large radius.
Using the same method, we estimate a radius
of $0.0429^{+0.0213}_{-0.0122} r_{\sun}$ and
a large uncertain distance of $913^{+282}_{-403}$\,pc
for the known ELM \object{SDSS J091709.55+463821.7}
\citepads{2016ApJ...818..155B}, which shows similar 
atmospheric parameters to \object{CA078}
($\log{g}\approx6$ and $T_{\rm eff}\approx$12\,000\,K).
The estimated radius
is about four times larger than the canonical WD radius
corresponding to an increase in magnitude of about 3\,mag, 
similar to what we see for \object{CA078} 
in Figs.~\ref{Fig_MG_G_J} and \ref{Fig_MG_BP_RP}.
While \object{SDSS J091709.55+463821.7} has only an
uncertain Gaia DR2 parallax and does not fall in the high-quality
Gaia DR2 100\,pc sample, we can compute its absolute $G$ magnitude 
as about 9.1\,mag, which
is only slighly larger than that of \object{CA078} ($\approx$8.3\,mag).
Both objects have almost exactly the same zero colour indices $BP-RP$
and would appear very close to each other in the Gaia DR2 CMD of
Fig.~\ref{Fig_MG_BP_RP}. Therefore, we conclude that \object{CA078}
is probably an ELM WD similar to \object{SDSS J091709.55+463821.7}
but at much smaller distance.

We have cross-identified the entire ELM survey sample of
\citetads{2016ApJ...818..155B} with Gaia DR2. Out of all 88 objects,
18 (with previous distance estimates between 0.42\,kpc 
and 7.77\,kpc) have negative or no parallaxes measured in Gaia DR2. 
Out of the remaining 70 ELM WDs, there are only 14 previously considered 
as relatively nearby, with distances between 0.1\,kpc and 0.3\,kpc 
according to \citetads{2016ApJ...818..155B}. However, as can be seen in 
Fig.~\ref{Fig_ELM}, six of them (shown in black) are shifted to much 
larger distances according to their Gaia DR2 parallaxes, including
the previously assumed nearest ELM WD \object{UCAC2 48250979}.
On the other hand, eight objects (shown in red) have only slightly 
larger Gaia DR2 distances compared to their previous distance estimates. 
The three nearest new ELM WDs are \object{LP 413-40},
\object{SDSS J075519.48+480034.0}, and \object{HS 1102+0934} at
distances of 182\,pc, 183\,pc, and 191\,pc.
From these three objects, only \object{LP 413-40} is included in the clean
sample of ELM WD binaries \citepads{2016ApJ...818..155B} and lies
with $BP-RP\approx+0.45$\,mag and an absolute G magnitude of 
$\approx$10.34\,mag almost 3\,mag above the normal WD sequence
in the Gaia CMD (cf. Fig.~\ref{Fig_MG_BP_RP}), similar to \object{CA078}. 
We conclude that \object{CA078}, which is located about three times closer
to the Sun and has a higher effective temperature than \object{LP 413-40},
represents a good new candidate for the nearest ELM WD.

\subsection{Outlook}
\label{Sect_outlook}

We have shown the Gaia DR2 colours $BP-RP$ to be a very effective tool 
for estimating effective temperatures of WDs. This can probably be
further enhanced when a larger sample of known WDs will be considered.
The polynomial fitting with our small sample of known WDs already
allowed us to reach uncertainties between $\pm$1100\,K and $\pm$300\,K for 
hot and cool WDs. So, we can e.g. 
for the recently discovered new WD member of the 
10\,pc sample, \object{TYC 3980-1081-1 B} \citepads{2018A&A...613A..26S},
estimate an effective 
temperature of 5100$\pm$300\,K
based on its Gaia DR2 colour of $BP-RP\approx1.06$\,mag.

We think that our new WDs, but also some of our rejected WDs,
deserve further attention with follow-up observations, including
higher-resolution spectroscopy,
radial velocity monitoring, variability analysis,
 and imaging (to search for possible
companions). In particular, the nature of the overluminous DA4.2$\pm$0.5 
ELM WD candidate \object{CA078} needs to be clarified.
If confirmed, this would be the nearest ELM WD. The
DB4.1$\pm$0.4
WD \object{CA103} can be studied in more detail as one of the nearest
representatives of its class. Among our rejected WDs, the G-type
carbon dwarf \object{CA171} is probably a promising target for
high-resolution spectroscopy,
as it is as bright ($G\approx13.35$\,mag) as the well-known 
and metal-poor ([Fe/H]$=-4$) dC star \object{G 77-61}. 
The latter is according to \citetads{2018RNAAS...2b..43M}
the only dC star with a detailed abundance analysis 
\citepads{2005A&A...434.1117P} so far.

\begin{acknowledgements}
Most of our spectroscopic observations were made within
a multi-season observing campaign carried out with the
2.2\,m telescope of the Centro Astron\'omico Hispano-Aleman 
at Calar Alto, Spain.
Part of these observations were carried out in service mode. 
We would like to thank the Calar Alto staff for their
kind support and for their help with the observations.
We thank Detlev Koester for providing his model spectra
of DA WDs,
Stephan Geier for helpful advice, 
and the anonymous referee for his kind report.
We have extensively used SIMBAD and VizieR at the CDS/Strasbourg.
\end{acknowledgements}

%
%

\bibliographystyle{aa}
\bibliography{spwdgref}
\end{document}